# Investigating the evolution of quantum entanglement of a qubit-qubit system with Dzyaloshinskii-Moriya interaction in the presence of magnetic fields

Seyed Mohsen Moosavi Khansari[1]
*Department of Physics, Faculty of Basic Sciences, Ayatollah Boroujerdi University, Boroujerd, IRAN*
Fazlollah Kazemi Hasanvand[2]
*Department of Physics, Faculty of Basic Sciences, Ayatollah Boroujerdi University, Boroujerd, IRAN*

In this paper, the quantum entanglement dynamics of a qubit-qubit compound system in the isotropic $XXX$ Heisenberg and anisotropic $XYZ$ models with $Dzyaloshinskii\text{-}Moriya$ interaction under magnetic fields is investigated. The system's initial state is considered as a spin coherence state, and the entanglement dynamics of this compound system is analyzed using the negativity criterion as an entanglement criterion to assess the impact of $Dzyaloshinskii\text{-}Moriya$ interaction and magnetic fields.

[1] E-mail of the corresponding author: m.moosavikhansari@abru.ac.ir
[2] E-mail: fa_kazemi270@yahoo.com

## 1 Introduction

Entanglement stands out as a remarkable outcome of quantum mechanics, pivotal in information theory and quantum computations [1- 5]. Coherent states, resembling classical states most closely [6, 7], possess the ability to form robust quantum correlations when combined [7, 8]. Studying the entanglement dynamics of qubit-qubit composite spin systems is deemed essential [9, 10]. This study delves into the entanglement dynamics of a qubit-qubit complex system featuring $Dzyaloshinskii$ -$Moriya(DM)$ interaction under the influence of magnetic fields [12-14]. Initially, a combination of spin-coherent states associated with spin $1/2$ is considered as the initial state of the qubit-qubit complex system. Subsequently, utilizing the negative criterion, we analyze the entanglement dynamics of the qubit-cube complex system [11]. The article will commence by outlining the isotropic Heisenberg models $XXX$ and then proceed to the anisotropic $XYZ$ model with magnetic fields affecting both qubits. The article's structure is outlined as follows: The initial portion of our exploration encompasses the in-depth exposition of the indispensable theoretical calculations crucial for the seamless execution of this scientific inquiry. Proceeding to the subsequent section, a detailed unveiling of the initial state governing the intricate qubit-qubit composite system alongside the transformative evolution orchestrated by the Hamiltonian highlighted in the inaugural section is thoroughly elucidated. Transitioning to the forthcoming section, a meticulous scrutiny is conducted to determine and quantify the negativity inherent in the requisite interactions. The subsequent segment is dedicated to the comprehensive revelation and assessment of the research outcomes, followed by an extensive discourse on the implications and interpretations ensuing from the findings. Lastly, the conclusive section synthesizes the research findings, encapsulating the key insights and implications delineating the culmination of this scholarly endeavor.

## 2 Theoretical calculations

We investigate the composite qubit-qubit system interacting with $DM$ , where each qubit is influenced by its individual external magnetic field. This analysis covers both the isotropic $XXX$ state and the anisotropic $XYZ$ state. The Hamiltonian's generic form is presented below:

$$H = \sum_k \frac{1}{4} J_k \sigma_a^k \sigma_b^k + \frac{1}{2}\left(\sigma_a^z B_{z,a} + \sigma_b^z B_{z,b}\right) + \frac{1}{4}\vec{D} \cdot \vec{\sigma}_a \times \vec{\sigma}_b \quad (1)$$

where $\mathrm{k} = \mathrm{x, y, z}$.

The extended relation of this Hamiltonian is as follows:

$$H = \frac{1}{2}\sigma_a^z B_{z,a} + \frac{1}{2}\sigma_b^z B_{z,b} - \frac{1}{4}D_x \sigma_a^z \sigma_b^y + \frac{1}{4}D_x \sigma_a^y \sigma_b^z + \frac{1}{4}D_y \sigma_a^z \sigma_b^x - \frac{1}{4}D_y \sigma_a^x \sigma_b^z - \frac{1}{4}D_z \sigma_a^y \sigma_b^x + \frac{1}{4}D_z \sigma_a^x \sigma_b^y + \frac{1}{4}J_x \sigma_a^x \sigma_b^x + \frac{1}{4}J_y \sigma_a^y \sigma_b^y + \frac{1}{4}J_z \sigma_a^z \sigma_b^z \quad (2)$$

where $B_{z,a}$ represents the external magnetic field acting on qubit $a$, and $B_{z,b}$ denotes the external magnetic field acting on qubit $b$. Both fields, denoted as $B_{z,a}$ and $B_{z,b}$, are oriented in the same direction, which is along the z-axis. This means that both fields share a common alignment with respect to the z-axis in the three-dimensional coordinate system. Magnetic fields are commonly measured in units known as Tesla, which is the standard unit of measurement used for quantifying their strength and intensity. $\sigma_i^x$, $\sigma_i^y$, and $\sigma_i^z$ where $i = a, b$ are defined as Pauli operators for qubits $a$ and $b$. $D_i$ for $i = x, y, z$ represent the components of the $DM$ interaction coefficient. $J_i$ with $i = x, y, z$ denote the magnitude of spin qubit-qubit interaction. For $J_x = J_y = J_z$, we obtain isotropic $XXX$ Heisenberg models, whereas for $J_x \neq J_y \neq J_z$, we have the $XYZ$ anisotropic model. Isotropy implies that a system exhibits the same properties in all directions. In the realm of physics, this means that physical properties, such as magnetic susceptibility or conductivity, are independent of the direction in which they are measured. We employ the negativity criterion for computing quantum entanglement.

Quantum negativity is a measure of quantum entanglement that quantifies the degree to which a quantum state deviates from being separable. It is particularly useful for assessing entanglement in mixed states. Here are some common methods for calculating quantum negativity:
The first step involves taking the partial transpose of the density matrix of the quantum state with respect to one of the subsystems. For a bipartite system described by a density matrix $\rho$, the partial transpose with respect to subsystem $A$ is denoted as $\rho^{T_A}$. Eigenvalue Calculation:
After obtaining the partially transposed density matrix, the next step is to compute its eigenvalues. The eigenvalues can be found using numerical methods or analytical techniques, depending on the complexity of the density matrix.
Negativity for a quantum state with density matrix $\rho$ is defined as follows [15-17]:
$N(\rho) = \frac{1}{2}\left(\left\|\rho^{T_i}\right\| - 1\right)$ (3)
within this framework, $\rho^{T_i}$ denotes the partial transpose of $\rho$ with regard to the specific component identified as $i$. If $N > 0$, the state is entangled and if $N = 0$, the state is separable. The larger the negativity, the stronger the entanglement. For complex systems, numerical techniques such as Monte Carlo methods or optimization algorithms may be employed to estimate the negativity, especially when dealing with large density matrices. These methods provide a systematic approach to quantifying quantum entanglement in various quantum systems, aiding in the understanding of quantum correlations.

## 3 The spin coherent state serves as the initial state of the qubit-qubit system

The spin coherent state is described as follows:
$|\alpha, j\rangle = (|\alpha|^2 + 1)^{-j} \times$
$\sum_{m=-j}^{j} \sqrt{\binom{2j}{j+m}} \alpha^{j+m} |j, m\rangle$ (4)
the qubit's coherent state is determined by setting $j = 1/2$ as follows:
$|\alpha, \frac{1}{2}\rangle = \frac{\left|-\frac{1}{2}, \frac{1}{2}\right\rangle}{\sqrt{|\alpha|^2+1}} + \frac{\alpha\left|\frac{1}{2}, \frac{1}{2}\right\rangle}{\sqrt{|\alpha|^2+1}}$ (5)
by utilizing the substitutions $\left|-\frac{1}{2}, \frac{1}{2}\right\rangle \to |0\rangle$ and $\left|\frac{1}{2}, \frac{1}{2}\right\rangle \to |1\rangle$, this state can be expressed as follows
$|\alpha, \frac{1}{2}\rangle = \frac{|0\rangle}{\sqrt{|\alpha|^2+1}} + \frac{\alpha|1\rangle}{\sqrt{|\alpha|^2+1}}$ (6)

We create a pure entangled state by superposing spin-coherent states associated with qubit $a$ and qubit $b$, as follows:
$$|\psi(0)\rangle = |00\rangle \left(\frac{\cos(\theta)}{\sqrt{\mathcal{N}}\sqrt{|\alpha_1|^2+1}\sqrt{|\beta_1|^2+1}} + \frac{e^{-i\varphi}\sin(\theta)}{\sqrt{\mathcal{N}}\sqrt{|\alpha_2|^2+1}\sqrt{|\beta_2|^2+1}}\right) + |01\rangle \left(\frac{\beta_1\cos(\theta)}{\sqrt{\mathcal{N}}\sqrt{|\alpha_1|^2+1}\sqrt{|\beta_1|^2+1}} + \frac{\beta_2 e^{-i\varphi}\sin(\theta)}{\sqrt{\mathcal{N}}\sqrt{|\alpha_2|^2+1}\sqrt{|\beta_2|^2+1}}\right) + |10\rangle \left(\frac{\alpha_1\cos(\theta)}{\sqrt{\mathcal{N}}\sqrt{|\alpha_1|^2+1}\sqrt{|\beta_1|^2+1}} + \frac{\alpha_2 e^{-i\varphi}\sin(\theta)}{\sqrt{\mathcal{N}}\sqrt{|\alpha_2|^2+1}\sqrt{|\beta_2|^2+1}}\right) + |11\rangle \left(\frac{\alpha_1\beta_1\cos(\theta)}{\sqrt{\mathcal{N}}\sqrt{|\alpha_1|^2+1}\sqrt{|\beta_1|^2+1}} + \frac{\alpha_2\beta_2 e^{-i\varphi}\sin(\theta)}{\sqrt{\mathcal{N}}\sqrt{|\alpha_2|^2+1}\sqrt{|\beta_2|^2+1}}\right) \quad (7)$$
where the variables $\theta$ and $\varphi$ are real numbers. By employing substitutions $\alpha_1 = \alpha_2 = \alpha$ and $\beta_1 = \beta_2 = -\alpha$, this quantum state can be represented as follows
$$|\psi(0)\rangle = |11\rangle \left(-\frac{\alpha^2\cos(\theta)}{\sqrt{\mathcal{N}}(|\alpha|^2+1)} - \frac{\alpha^2 e^{-i\varphi}\sin(\theta)}{\sqrt{\mathcal{N}}(|\alpha|^2+1)}\right) + |00\rangle \left(\frac{\cos(\theta)}{\sqrt{\mathcal{N}}(|\alpha|^2+1)} + \frac{e^{-i\varphi}\sin(\theta)}{\sqrt{\mathcal{N}}(|\alpha|^2+1)}\right) + |01\rangle \left(-\frac{\alpha\cos(\theta)}{\sqrt{\mathcal{N}}(|\alpha|^2+1)} - \frac{\alpha e^{-i\varphi}\sin(\theta)}{\sqrt{\mathcal{N}}(|\alpha|^2+1)}\right) + |10\rangle \left(\frac{\alpha\cos(\theta)}{\sqrt{\mathcal{N}}(|\alpha|^2+1)} + \frac{\alpha e^{-i\varphi}\sin(\theta)}{\sqrt{\mathcal{N}}(|\alpha|^2+1)}\right) \quad (8)$$
The normalization relation of this quantum state is determined as follows
$$\langle\psi(0)|\psi(0)\rangle = -\frac{\alpha^2\cos(\theta)\left(-\frac{\cos(\theta)\alpha^2}{\sqrt{\mathcal{N}}(|\alpha|^2+1)} - \frac{e^{-i\varphi}\sin(\theta)\alpha^2}{\sqrt{\mathcal{N}}(|\alpha|^2+1)}\right)^*}{\sqrt{\mathcal{N}}(|\alpha|^2+1)} - \frac{\alpha^2 e^{-i\varphi}\sin(\theta)\left(-\frac{\cos(\theta)\alpha^2}{\sqrt{\mathcal{N}}(|\alpha|^2+1)} - \frac{e^{-i\varphi}\sin(\theta)\alpha^2}{\sqrt{\mathcal{N}}(|\alpha|^2+1)}\right)^*}{\sqrt{\mathcal{N}}(|\alpha|^2+1)} -$$

$$\frac{\alpha\cos(\theta)\left(-\frac{\alpha\cos(\theta)}{\sqrt{\mathcal{N}}(|\alpha|^2+1)}-\frac{e^{-i\varphi}\alpha\sin(\theta)}{\sqrt{\mathcal{N}}(|\alpha|^2+1)}\right)^*}{\sqrt{\mathcal{N}}(|\alpha|^2+1)}+\frac{\alpha\cos(\theta)\left(\frac{\alpha\cos(\theta)}{\sqrt{\mathcal{N}}(|\alpha|^2+1)}+\frac{e^{-i\varphi}\alpha\sin(\theta)}{\sqrt{\mathcal{N}}(|\alpha|^2+1)}\right)^*}{\sqrt{\mathcal{N}}(|\alpha|^2+1)}$$

$$-\frac{\alpha e^{-i\varphi}\sin(\theta)\left(-\frac{\alpha\cos(\theta)}{\sqrt{\mathcal{N}}(|\alpha|^2+1)}-\frac{e^{-i\varphi}\alpha\sin(\theta)}{\sqrt{\mathcal{N}}(|\alpha|^2+1)}\right)^*}{\sqrt{\mathcal{N}}(|\alpha|^2+1)}+\frac{\alpha e^{-i\varphi}\sin(\theta)\left(\frac{\alpha\cos(\theta)}{\sqrt{\mathcal{N}}(|\alpha|^2+1)}+\frac{e^{-i\varphi}\alpha\sin(\theta)}{\sqrt{\mathcal{N}}(|\alpha|^2+1)}\right)^*}{\sqrt{\mathcal{N}}(|\alpha|^2+1)}$$

$$+\frac{\cos(\theta)\left(\frac{\cos(\theta)}{\sqrt{\mathcal{N}}(|\alpha|^2+1)}+\frac{e^{-i\varphi}\sin(\theta)}{\sqrt{\mathcal{N}}(|\alpha|^2+1)}\right)^*}{\sqrt{\mathcal{N}}(|\alpha|^2+1)}+\frac{e^{-i\varphi}\sin(\theta)\left(\frac{\cos(\theta)}{\sqrt{\mathcal{N}}(|\alpha|^2+1)}+\frac{e^{-i\varphi}\sin(\theta)}{\sqrt{\mathcal{N}}(|\alpha|^2+1)}\right)^*}{\sqrt{\mathcal{N}}(|\alpha|^2+1)} \quad (9)$$

Presently, the density operator is capable of being represented in the Dirac notation as illustrated below

$$\rho(0) = -\frac{\left(\frac{\cos(\theta)}{\sqrt{\mathcal{N}}(|\alpha|^2+1)}+\frac{e^{-i\varphi}\sin(\theta)}{\sqrt{\mathcal{N}}(|\alpha|^2+1)}\right)^*\cos(\theta)|11\rangle\langle 00|\alpha^2}{\sqrt{\mathcal{N}}(|\alpha|^2+1)}-\frac{e^{-i\varphi}\left(\frac{\cos(\theta)}{\sqrt{\mathcal{N}}(|\alpha|^2+1)}+\frac{e^{-i\varphi}\sin(\theta)}{\sqrt{\mathcal{N}}(|\alpha|^2+1)}\right)^*\sin(\theta)|11\rangle\langle 00|\alpha^2}{\sqrt{\mathcal{N}}(|\alpha|^2+1)}$$

$$-\frac{\left(-\frac{\alpha\cos(\theta)}{\sqrt{\mathcal{N}}(|\alpha|^2+1)}-\frac{e^{-i\varphi}\alpha\sin(\theta)}{\sqrt{\mathcal{N}}(|\alpha|^2+1)}\right)^*\cos(\theta)|11\rangle\langle 01|\alpha^2}{\sqrt{\mathcal{N}}(|\alpha|^2+1)}$$

$$-\frac{\left(\frac{\alpha\cos(\theta)}{\sqrt{\mathcal{N}}(|\alpha|^2+1)}+\frac{e^{-i\varphi}\alpha\sin(\theta)}{\sqrt{\mathcal{N}}(|\alpha|^2+1)}\right)^*\cos(\theta)|11\rangle\langle 10|\alpha^2}{\sqrt{\mathcal{N}}(|\alpha|^2+1)}-\frac{e^{-i\varphi}\left(\frac{\alpha\cos(\theta)}{\sqrt{\mathcal{N}}(|\alpha|^2+1)}+\frac{e^{-i\varphi}\alpha\sin(\theta)}{\sqrt{\mathcal{N}}(|\alpha|^2+1)}\right)^*\sin(\theta)|11\rangle\langle 10|\alpha^2}{\sqrt{\mathcal{N}}(|\alpha|^2+1)}$$

$$-\frac{\left(-\frac{\cos(\theta)\alpha^2}{\sqrt{\mathcal{N}}(|\alpha|^2+1)}-\frac{e^{-i\varphi}\sin(\theta)\alpha^2}{\sqrt{\mathcal{N}}(|\alpha|^2+1)}\right)^*\cos(\theta)|11\rangle\langle 11|\alpha^2}{\sqrt{\mathcal{N}}(|\alpha|^2+1)}-\frac{e^{-i\varphi}\left(-\frac{\cos(\theta)\alpha^2}{\sqrt{\mathcal{N}}(|\alpha|^2+1)}-\frac{e^{-i\varphi}\sin(\theta)\alpha^2}{\sqrt{\mathcal{N}}(|\alpha|^2+1)}\right)^*\sin(\theta)|11\rangle\langle 11|\alpha^2}{\sqrt{\mathcal{N}}(|\alpha|^2+1)}$$

$$-\frac{\left(\frac{\cos(\theta)}{\sqrt{\mathcal{N}}(|\alpha|^2+1)}+\frac{e^{-i\varphi}\sin(\theta)}{\sqrt{\mathcal{N}}(|\alpha|^2+1)}\right)^*\cos(\theta)|01\rangle\langle 00|\alpha}{\sqrt{\mathcal{N}}(|\alpha|^2+1)}-\frac{e^{-i\varphi}\left(\frac{\cos(\theta)}{\sqrt{\mathcal{N}}(|\alpha|^2+1)}+\frac{e^{-i\varphi}\sin(\theta)}{\sqrt{\mathcal{N}}(|\alpha|^2+1)}\right)^*\sin(\theta)|01\rangle\langle 00|\alpha}{\sqrt{\mathcal{N}}(|\alpha|^2+1)}$$

$$+\frac{\left(\frac{\cos(\theta)}{\sqrt{\mathcal{N}}(|\alpha|^2+1)}+\frac{e^{-i\varphi}\sin(\theta)}{\sqrt{\mathcal{N}}(|\alpha|^2+1)}\right)^*\cos(\theta)|10\rangle\langle 00|\alpha}{\sqrt{\mathcal{N}}(|\alpha|^2+1)}+\frac{e^{-i\varphi}\left(\frac{\cos(\theta)}{\sqrt{\mathcal{N}}(|\alpha|^2+1)}+\frac{e^{-i\varphi}\sin(\theta)}{\sqrt{\mathcal{N}}(|\alpha|^2+1)}\right)^*\sin(\theta)|10\rangle\langle 00|\alpha}{\sqrt{\mathcal{N}}(|\alpha|^2+1)}$$

$$-\frac{\left(-\frac{\alpha\cos(\theta)}{\sqrt{\mathcal{N}}(|\alpha|^2+1)}-\frac{e^{-i\varphi}\alpha\sin(\theta)}{\sqrt{\mathcal{N}}(|\alpha|^2+1)}\right)^*\cos(\theta)|01\rangle\langle 01|\alpha}{\sqrt{\mathcal{N}}(|\alpha|^2+1)}-\frac{e^{-i\varphi}\left(-\frac{\alpha\cos(\theta)}{\sqrt{\mathcal{N}}(|\alpha|^2+1)}-\frac{e^{-i\varphi}\alpha\sin(\theta)}{\sqrt{\mathcal{N}}(|\alpha|^2+1)}\right)^*\sin(\theta)|01\rangle\langle 01|\alpha}{\sqrt{\mathcal{N}}(|\alpha|^2+1)}$$

$$+\frac{\left(-\frac{\alpha\cos(\theta)}{\sqrt{\mathcal{N}}(|\alpha|^2+1)}-\frac{e^{-i\varphi}\alpha\sin(\theta)}{\sqrt{\mathcal{N}}(|\alpha|^2+1)}\right)^*\cos(\theta)|10\rangle\langle 01|\alpha}{\sqrt{\mathcal{N}}(|\alpha|^2+1)}+\frac{e^{-i\varphi}\left(-\frac{\alpha\cos(\theta)}{\sqrt{\mathcal{N}}(|\alpha|^2+1)}-\frac{e^{-i\varphi}\alpha\sin(\theta)}{\sqrt{\mathcal{N}}(|\alpha|^2+1)}\right)^*\sin(\theta)|10\rangle\langle 01|\alpha}{\sqrt{\mathcal{N}}(|\alpha|^2+1)}$$

$$\frac{\left(\frac{\alpha\cos(\theta)}{\sqrt{\mathcal{N}}(|\alpha|^2+1)}+\frac{e^{-i\varphi}\alpha\sin(\theta)}{\sqrt{\mathcal{N}}(|\alpha|^2+1)}\right)^*\cos(\theta)|01\rangle\langle 10|\alpha}{\sqrt{\mathcal{N}}(|\alpha|^2+1)}-\frac{e^{-i\varphi}\left(\frac{\alpha\cos(\theta)}{\sqrt{\mathcal{N}}(|\alpha|^2+1)}+\frac{e^{-i\varphi}\alpha\sin(\theta)}{\sqrt{\mathcal{N}}(|\alpha|^2+1)}\right)^*\sin(\theta)|01\rangle\langle 10|\alpha}{\sqrt{\mathcal{N}}(|\alpha|^2+1)}$$

$$+$$

$$\frac{\left(\frac{\alpha\cos(\theta)}{\sqrt{\mathcal{N}}(|\alpha|^2+1)}+\frac{e^{-i\varphi}\alpha\sin(\theta)}{\sqrt{\mathcal{N}}(|\alpha|^2+1)}\right)^*\cos(\theta)|10\rangle\langle 10|\alpha}{\sqrt{\mathcal{N}}(|\alpha|^2+1)}$$

$$+\frac{e^{-i\varphi}\left(\frac{\alpha\cos(\theta)}{\sqrt{\mathcal{N}}(|\alpha|^2+1)}+\frac{e^{-i\varphi}\alpha\sin(\theta)}{\sqrt{\mathcal{N}}(|\alpha|^2+1)}\right)^*\sin(\theta)|10\rangle\langle 10|\alpha}{\sqrt{\mathcal{N}}(|\alpha|^2+1)}$$

$$-\frac{\left(-\frac{\cos(\theta)\alpha^2}{\sqrt{\mathcal{N}}(|\alpha|^2+1)}-\frac{e^{-i\varphi}\sin(\theta)\alpha^2}{\sqrt{\mathcal{N}}(|\alpha|^2+1)}\right)^*\cos(\theta)|01\rangle\langle 11|\alpha}{\sqrt{\mathcal{N}}(|\alpha|^2+1)}$$

$$-\frac{e^{-i\varphi}\left(-\frac{\cos(\theta)\alpha^2}{\sqrt{\mathcal{N}}(|\alpha|^2+1)}-\frac{e^{-i\varphi}\sin(\theta)\alpha^2}{\sqrt{\mathcal{N}}(|\alpha|^2+1)}\right)^*\sin(\theta)|01\rangle\langle 11|}{\sqrt{\mathcal{N}}(|\alpha|^2+1)}$$

$$+\frac{\left(-\frac{\cos(\theta)\alpha^2}{\sqrt{\mathcal{N}}(|\alpha|^2+1)}-\frac{e^{-i\varphi}\sin(\theta)\alpha^2}{\sqrt{\mathcal{N}}(|\alpha|^2+1)}\right)^*\cos(\theta)|10\rangle\langle 11|\alpha}{\sqrt{\mathcal{N}}(|\alpha|^2+1)}$$

$$+\frac{e^{-i\varphi}\left(-\frac{\cos(\theta)\alpha^2}{\sqrt{\mathcal{N}}(|\alpha|^2+1)}-\frac{e^{-i\varphi}\sin(\theta)\alpha^2}{\sqrt{\mathcal{N}}(|\alpha|^2+1)}\right)^*\sin(\theta)|10\rangle\langle 11|}{\sqrt{\mathcal{N}}(|\alpha|^2+1)}$$

$$+\frac{\left(\frac{\cos(\theta)}{\sqrt{\mathcal{N}}(|\alpha|^2+1)}+\frac{e^{-i\varphi}\sin(\theta)}{\sqrt{\mathcal{N}}(|\alpha|^2+1)}\right)^*\cos(\theta)|00\rangle\langle 00|}{\sqrt{\mathcal{N}}(|\alpha|^2+1)}$$

$$+\frac{e^{-i\varphi}\left(\frac{\cos(\theta)}{\sqrt{\mathcal{N}}(|\alpha|^2+1)}+\frac{e^{-i\varphi}\sin(\theta)}{\sqrt{\mathcal{N}}(|\alpha|^2+1)}\right)^*\sin(\theta)|00\rangle\langle 00|}{\sqrt{\mathcal{N}}(|\alpha|^2+1)}$$

$$+\frac{\left(-\frac{\alpha\cos(\theta)}{\sqrt{\mathcal{N}}(|\alpha|^2+1)}-\frac{e^{-i\varphi}\alpha\sin(\theta)}{\sqrt{\mathcal{N}}(|\alpha|^2+1)}\right)^*\cos(\theta)|00\rangle\langle 01|}{\sqrt{\mathcal{N}}(|\alpha|^2+1)}$$

$$+\frac{e^{-i\varphi}\left(-\frac{\alpha\cos(\theta)}{\sqrt{\mathcal{N}}(|\alpha|^2+1)}-\frac{e^{-i\varphi}\alpha\sin(\theta)}{\sqrt{\mathcal{N}}(|\alpha|^2+1)}\right)^*\sin(\theta)|00\rangle\langle 01|}{\sqrt{\mathcal{N}}(|\alpha|^2+1)}$$

$$+\frac{\left(\frac{\alpha\cos(\theta)}{\sqrt{\mathcal{N}}(|\alpha|^2+1)}+\frac{e^{-i\varphi}\alpha\sin(\theta)}{\sqrt{\mathcal{N}}(|\alpha|^2+1)}\right)^*\cos(\theta)|00\rangle\langle 10|}{\sqrt{\mathcal{N}}(|\alpha|^2+1)}$$

$$+\frac{e^{-i\varphi}\left(\frac{\alpha\cos(\theta)}{\sqrt{\mathcal{N}}(|\alpha|^2+1)}+\frac{e^{-i\varphi}\alpha\sin(\theta)}{\sqrt{\mathcal{N}}(|\alpha|^2+1)}\right)^*\sin(\theta)|00\rangle\langle 10|}{\sqrt{\mathcal{N}}(|\alpha|^2+1)}$$

$$+\frac{\left(-\frac{\cos(\theta)\alpha^2}{\sqrt{\mathcal{N}}(|\alpha|^2+1)}-\frac{e^{-i\varphi}\sin(\theta)\alpha^2}{\sqrt{\mathcal{N}}(|\alpha|^2+1)}\right)^*\cos(\theta)|00\rangle\langle 11|}{\sqrt{\mathcal{N}}(|\alpha|^2+1)}+\frac{e^{-i\varphi}\left(-\frac{\cos(\theta)\alpha^2}{\sqrt{\mathcal{N}}(|\alpha|^2+1)}-\frac{e^{-i\varphi}\sin(\theta)\alpha^2}{\sqrt{\mathcal{N}}(|\alpha|^2+1)}\right)^*\sin(\theta)|00\rangle\langle 11|}{\sqrt{\mathcal{N}}(|\alpha|^2+1)} \quad (10)$$

Using this quantum state, the initial state density matrix arrays of the system can be computed as follows

$$\rho_{1,1}(0)=\frac{e^{-i\varphi}(\sin(\theta)+e^{i\varphi}\cos(\theta))\left(\frac{\cos(\theta)+e^{-i\varphi}\sin(\theta)}{\sqrt{\mathcal{N}}}\right)^*}{\sqrt{\mathcal{N}}(|\alpha|^2+1)^2} \quad (11)$$

$$\rho_{1,2}(0)=-\frac{e^{-i(\varphi-\varphi^*)}(\sin(\theta)+e^{i\varphi}\cos(\theta))\left(\frac{\alpha\left(e^{i\varphi}\cos(\theta)+\sin(\theta)\right)}{\sqrt{\mathcal{N}}}\right)^*}{\sqrt{\mathcal{N}}(|\alpha|^2+1)^2} \quad (12)$$

$$\rho_{1,3}(0)=\frac{e^{-i(\varphi-\varphi^*)}(\sin(\theta)+e^{i\varphi}\cos(\theta))\left(\frac{\alpha\left(e^{i\varphi}\cos(\theta)+\sin(\theta)\right)}{\sqrt{\mathcal{N}}}\right)^*}{\sqrt{\mathcal{N}}(|\alpha|^2+1)^2} \quad (13)$$

$$\rho_{1,4}(0)=-\frac{(\alpha^*)^2e^{-i(\varphi-\varphi^*)}(\sin(\theta)+e^{i\varphi}\cos(\theta))\left(\frac{e^{i\varphi}\cos(\theta)+\sin(\theta)}{\sqrt{\mathcal{N}}}\right)^*}{\sqrt{\mathcal{N}}(|\alpha|^2+1)^2} \quad (14)$$

$$\rho_{2,1}(0)=-\frac{\alpha e^{-i\varphi}(\sin(\theta)+e^{i\varphi}\cos(\theta))\left(\frac{\cos(\theta)+e^{-i\varphi}\sin(\theta)}{\sqrt{\mathcal{N}}}\right)^*}{\sqrt{\mathcal{N}}(|\alpha|^2+1)^2} \quad (15)$$

$$\rho_{2,2}(0)=\frac{\alpha e^{-i(\varphi-\varphi^*)}(\sin(\theta)+e^{i\varphi}\cos(\theta))\left(\frac{\alpha\left(e^{i\varphi}\cos(\theta)+\sin(\theta)\right)}{\sqrt{\mathcal{N}}}\right)^*}{\sqrt{\mathcal{N}}(|\alpha|^2+1)^2} \quad (16)$$

$$\rho_{2,3}(0)=-\frac{\alpha e^{-i(\varphi-\varphi^*)}(\sin(\theta)+e^{i\varphi}\cos(\theta))\left(\frac{\alpha\left(e^{i\varphi}\cos(\theta)+\sin(\theta)\right)}{\sqrt{\mathcal{N}}}\right)^*}{\sqrt{\mathcal{N}}(|\alpha|^2+1)^2} \quad (17)$$

$$\rho_{2,4}(0)=\frac{\alpha(\alpha^*)^2e^{-i(\varphi-\varphi^*)}(\sin(\theta)+e^{i\varphi}\cos(\theta))\left(\frac{e^{i\varphi}\cos(\theta)+\sin(\theta)}{\sqrt{\mathcal{N}}}\right)^*}{\sqrt{\mathcal{N}}(|\alpha|^2+1)^2} \quad (18)$$

$$\rho_{3,1}(0)=\frac{\alpha e^{-i\varphi}(\sin(\theta)+e^{i\varphi}\cos(\theta))\left(\frac{\cos(\theta)+e^{-i\varphi}\sin(\theta)}{\sqrt{\mathcal{N}}}\right)^*}{\sqrt{\mathcal{N}}(|\alpha|^2+1)^2} \quad (19)$$

$$\rho_{3,2}(0) = -\frac{\alpha e^{-i(\varphi-\varphi^{*})}(\sin(\theta)+e^{i\varphi}\cos(\theta))\left(\frac{\alpha\left(e^{i\varphi}\cos(\theta)+\sin(\theta)\right)}{\sqrt{\mathcal{N}}}\right)^{*}}{\sqrt{\mathcal{N}}(|\alpha|^{2}+1)^{2}} \quad (20)$$

$$\rho_{3,3}(0) = \frac{\alpha e^{-i(\varphi-\varphi^{*})}(\sin(\theta)+e^{i\varphi}\cos(\theta))\left(\frac{\alpha\left(e^{i\varphi}\cos(\theta)+\sin(\theta)\right)}{\sqrt{\mathcal{N}}}\right)^{*}}{\sqrt{\mathcal{N}}(|\alpha|^{2}+1)^{2}} \quad (21)$$

$$\rho_{3,4}(0) = -\frac{\alpha(\alpha^{*})^{2}e^{-i(\varphi-\varphi^{*})}(\sin(\theta)+e^{i\varphi}\cos(\theta))\left(\frac{e^{i\varphi}\cos(\theta)+\sin(\theta)}{\sqrt{\mathcal{N}}}\right)^{*}}{\sqrt{\mathcal{N}}(|\alpha|^{2}+1)^{2}} \quad (22)$$

$$\rho_{4,1}(0) = -\frac{\alpha^{2}e^{-i\varphi}(\sin(\theta)+e^{i\varphi}\cos(\theta))\left(\frac{\cos(\theta)+e^{-i\varphi}\sin(\theta)}{\sqrt{\mathcal{N}}}\right)^{*}}{\sqrt{\mathcal{N}}(|\alpha|^{2}+1)^{2}} \quad (23)$$

$$\rho_{4,2}(0) = \frac{\alpha^{2}e^{-i(\varphi-\varphi^{*})}(\sin(\theta)+e^{i\varphi}\cos(\theta))\left(\frac{\alpha\left(e^{i\varphi}\cos(\theta)+\sin(\theta)\right)}{\sqrt{\mathcal{N}}}\right)^{*}}{\sqrt{\mathcal{N}}(|\alpha|^{2}+1)^{2}} \quad (24)$$

$$\rho_{4,3}(0) = -\frac{\alpha^{2}e^{-i(\varphi-\varphi^{*})}(\sin(\theta)+e^{i\varphi}\cos(\theta))\left(\frac{\alpha\left(e^{i\varphi}\cos(\theta)+\sin(\theta)\right)}{\sqrt{\mathcal{N}}}\right)^{*}}{\sqrt{\mathcal{N}}(|\alpha|^{2}+1)^{2}} \quad (25)$$

$$\rho_{4,4}(0) = \frac{\alpha^{2}(\alpha^{*})^{2}e^{-i(\varphi-\varphi^{*})}(\sin(\theta)+e^{i\varphi}\cos(\theta))\left(\frac{e^{i\varphi}\cos(\theta)+\sin(\theta)}{\sqrt{\mathcal{N}}}\right)^{*}}{\sqrt{\mathcal{N}}(|\alpha|^{2}+1)^{2}} \quad (26)$$

The Hamiltonian matrix for these two qubits can be determined in their respective bases as follows

$$H = \begin{pmatrix} \frac{1}{4}\left(J_z + 2\left(B_{z,a}+B_{z,b}\right)\right) & \frac{1}{4}\left(iD_x + D_y\right) & -\frac{1}{4}i\left(D_x - iD_y\right) & \frac{1}{4}\left(J_x - J_y\right) \\ \frac{1}{4}\left(D_y - iD_x\right) & \frac{1}{4}\left(-J_z + 2B_{z,a} - 2B_{z,b}\right) & \frac{1}{4}\left(2iD_z + J_x + J_y\right) & \frac{1}{4}\left(iD_x + D_y\right) \\ \frac{1}{4}i\left(D_x + iD_y\right) & \frac{1}{4}\left(-2iD_z + J_x + J_y\right) & \frac{1}{4}\left(-J_z - 2B_{z,a} + 2B_{z,b}\right) & -\frac{1}{4}i\left(D_x - iD_y\right) \\ \frac{1}{4}\left(J_x - J_y\right) & \frac{1}{4}\left(D_y - iD_x\right) & \frac{1}{4}i\left(D_x + iD_y\right) & \frac{1}{4}\left(J_z - 2\left(B_{z,a}+B_{z,b}\right)\right) \end{pmatrix} \quad (27)$$

Using this Hamiltonian, the evolution of the initial state $|\psi(0)\rangle$ after the effect of $U(t) = Exp(-iHt)$, can be expressed as $|\psi(t)\rangle = U(t)|\psi(0)\rangle$. The time-dependent density operator of the system can now be expressed as $\rho(t) = |\psi(t)\rangle\langle\psi(t)|$.

The anisotropic Heisenberg model is fundamental in the study of magnetic systems, where the complexities of spin interactions display directional dependence. This model expands upon the classical Heisenberg framework, permitting unique exchange interactions along various axes, thereby offering a more intricate landscape to investigate phenomena such as magnetic anisotropy, phase transitions, and critical behavior. In its most fundamental form, the Hamiltonian of the anisotropic Heisenberg model can be expressed as:

$$H = -J_x \sum_i S_i^x S_j^x - J_y \sum_i S_i^y S_j^y - J_z \sum_i S_i^z S_j^z \quad (28)$$

where $J_x$, $J_y$, and $J_z$ denote the exchange constants along the respective axes, and $S_i$ represents the spin operators at site i. The competition between these anisotropic terms gives rise to many magnetic states, with implications for both classical and quantum regimes. As one delves deeper into this model, it becomes evident that the interplay between the anisotropic interactions can lead to phenomena such as spin waves, magnetic ordering, and even non-trivial topological states. Depending on the relative magnitudes of $J_x$, $J_y$, and $J_z$, the system can exhibit various behaviours, from simple ferromagnetism to complex spin configurations that challenge our understanding of fundamental magnetic principles. Moreover, the anisotropic Heisenberg model is a valuable framework for exploring the effects of external fields, temperature variations, and disorder. Through various computational techniques ”from exact diagonalization to Monte Carlo simulations

”researchers can probe the phase diagrams and identify regions of stability, critical points, and the nature of excitations that emerge from this rich landscape. In the context of real materials, many systems exhibit anisotropic exchange interactions due to underlying crystal symmetries, spin-orbit coupling, or geometric frustration. Consequently, the anisotropic Heisenberg model not only provides theoretical insight but also paves the way for experimental investigations, wherein phenomena such as spin reorientation transitions or the emergence of magnetic skyrmions can be observed and understood. Thus, the anisotropic Heisenberg model stands as a profound illustration of how simple interactions can lead to emergent complexities, illuminating pathways for future research in condensed matter physics, material science, and beyond. Some common types of anisotropy include:

$XY$ model: Interactions between spins in the $x-y$ plane only.

$XXZ$ model: Anisotropy between the $x-y$ plane and the $z$-axis.

$XYZ$ model: Full anisotropy with different interactions along $x$, $y$, and $z$ directions. This research employs the $XYZ$ anisotropic Heisenberg model as a fundamental framework for our analysis. By utilizing this model, we aim to explore the intricate behaviors and interactions of spins within the system. The anisotropic Heisenberg model in the XYZ form is a theoretical framework used in condensed matter physics to describe magnetic interactions in a system of spins, where the interactions vary depending on the direction of the spins. Anisotropic: The term anisotropic indicates that the interactions between spins are direction-dependent. This means that the strength of the magnetic interactions can differ along different axes $(x, y, z)$. This model is used to study various phenomena in magnetic materials, including magnetic anisotropy, phase transitions, and the behavior of spin systems under external fields. It helps in understanding how different orientations of spins can lead to diverse magnetic properties in real materials.

## 4 Results and discussion

Within this segment, we will showcase the results that have been acquired through our research endeavors, followed by a comprehensive and detailed analysis of these findings. Numerical methodologies and computational approaches have been consciously incorporated and utilized to accurately compute and analyze every aspect of the results in the study.

This research specifically concentrates on a magnetic field intensity range that spans from 1 to 5 units. This focus is particularly noteworthy given that the intensity of the magnetic field can theoretically take on an unlimited number of values beyond this specified range. In this research, the range of values that are being considered for $D_i$ and $J_i$ for $i = x, y, z$ is specifically defined to be between $[0,1]$ and $[1,3]$, respectively. This careful selection of ranges is important for the analysis being conducted. In this research, we are specifically examining the millisecond time scale. Generally, this timescale is used to study the dynamics and behaviour of quantum systems, especially in quantum computing and information.

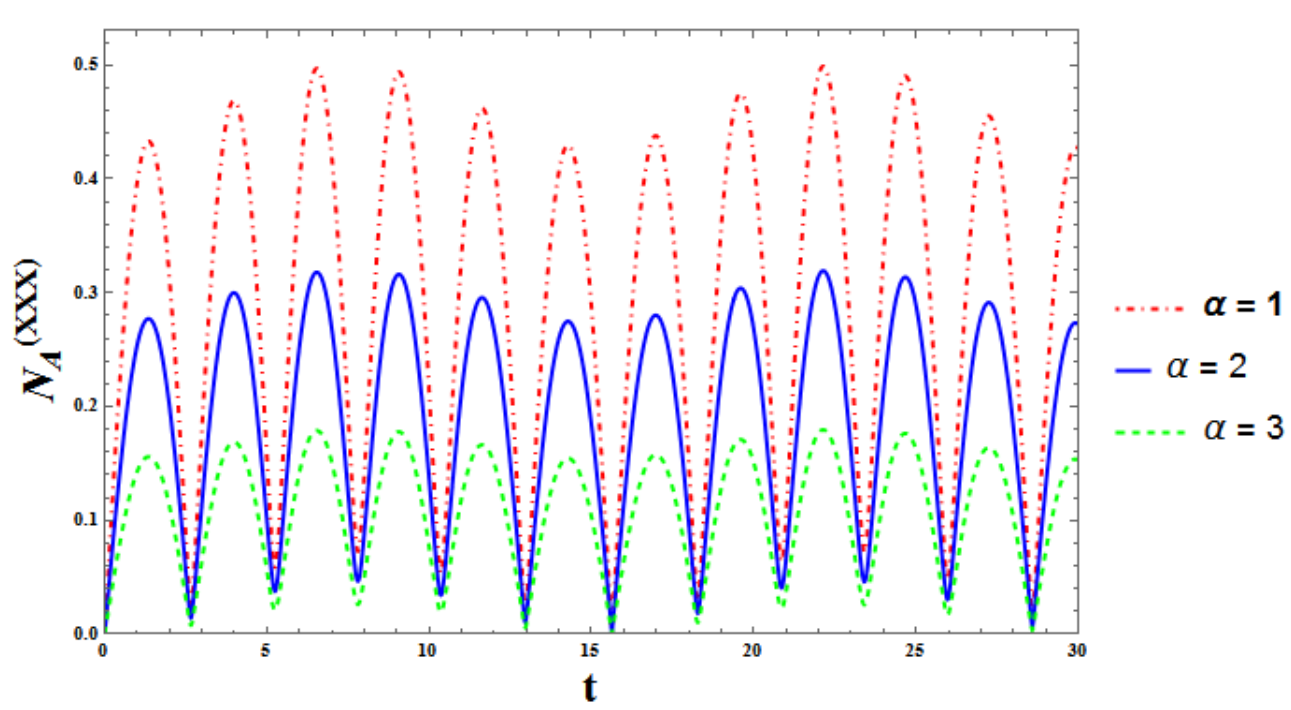

Figure 1: Time negativity diagram for the isotropic state $XXX$ $(J_x = J_y = J_z)$ of subsystem (qubit) $a$ is illustrated for three varying values of $\alpha$, with initial values of $J_x = 1, J_y = 1, J_z = 1, B_{z,a} = 1, B_{z,b} = 1, D_x = 0, D_y = 0, D_z = 1, \mathcal{N} = 2, \varphi = 0, \theta = \pi/4$. The magnetic fields applied to both qubits are equal $(B_{z,a} = B_{z,b})$ and oriented in the z-axis direction.

In Fig. 1, depicting the isotropic state of $XXX$, the negativity diagram for subsystem $a$ is illustrated when equal external fields are applied to both qubits. This graph reveals consistent fluctuations in negativity, or its entanglement equivalent, across all three $\alpha$ values. The oscillation period remains constant for all $\alpha$ values. The maximum fluctuations exhibit the same periodicity for all three cases. With an increase in $\alpha$, both maximum and minimum fluctuations decrease. No entanglement decay is observed in these three instances. The time-averaged entanglement is highest for $\alpha = 1$ (0.30), followed by $\alpha = 2$ with a time-averaged entanglement value of 0.192. The lowest time average is associated with $\alpha = 3$, at 0.108. These findings were derived from the following relationships

$$U(t) = \begin{pmatrix} e^{-\frac{5it}{4}} & 0 & 0 & 0 \\ 0 & \frac{1}{2}e^{\frac{it}{4}-\frac{it}{\sqrt{2}}} + \frac{1}{2}e^{\frac{it}{\sqrt{2}}+\frac{it}{4}} & \frac{\left(\frac{1}{2}+\frac{i}{2}\right)e^{\frac{it}{4}-\frac{it}{\sqrt{2}}}}{\sqrt{2}} - \frac{\left(\frac{1}{2}+\frac{i}{2}\right)e^{\frac{it}{\sqrt{2}}+\frac{it}{4}}}{\sqrt{2}} & 0 \\ 0 & \frac{\left(\frac{1}{2}-\frac{i}{2}\right)e^{\frac{it}{4}-\frac{it}{\sqrt{2}}}}{\sqrt{2}} - \frac{\left(\frac{1}{2}-\frac{i}{2}\right)e^{\frac{it}{\sqrt{2}}+\frac{it}{4}}}{\sqrt{2}} & \frac{1}{2}e^{\frac{it}{4}-\frac{it}{\sqrt{2}}} + \frac{1}{2}e^{\frac{it}{\sqrt{2}}+\frac{it}{4}} & 0 \\ 0 & 0 & 0 & e^{\frac{3it}{4}} \end{pmatrix} \quad (29)$$

and

$$|\psi(t)\rangle = U(t)|\psi(0)\rangle = -\frac{\alpha^2 e^{\frac{3it}{4}}|11\rangle}{|\alpha|^2+1} + \frac{e^{-\frac{5it}{4}}|00\rangle}{|\alpha|^2+1} + |01\rangle\left(\frac{\left(\frac{1}{2}+\frac{i}{2}\right)\alpha e^{\frac{it}{4}-\frac{it}{\sqrt{2}}}}{\sqrt{2}(|\alpha|^2+1)} - \frac{\alpha e^{\frac{it}{4}-\frac{it}{\sqrt{2}}}}{2(|\alpha|^2+1)} - \frac{\left(\frac{1}{2}+\frac{i}{2}\right)\alpha e^{\frac{it}{\sqrt{2}}+\frac{it}{4}}}{\sqrt{2}(|\alpha|^2+1)} - \frac{\alpha e^{\frac{it}{\sqrt{2}}+\frac{it}{4}}}{2(|\alpha|^2+1)}\right) + |10\rangle\left(-\frac{\left(\frac{1}{2}-\frac{i}{2}\right)\alpha e^{\frac{it}{4}-\frac{it}{\sqrt{2}}}}{\sqrt{2}(|\alpha|^2+1)} + \frac{\alpha e^{\frac{it}{4}-\frac{it}{\sqrt{2}}}}{2(|\alpha|^2+1)} + \frac{\left(\frac{1}{2}-\frac{i}{2}\right)\alpha e^{\frac{it}{\sqrt{2}}+\frac{it}{4}}}{\sqrt{2}(|\alpha|^2+1)} + \frac{\alpha e^{\frac{it}{\sqrt{2}}+\frac{it}{4}}}{2(|\alpha|^2+1)}\right) \quad (30)$$

and in the following, we will use the $\rho(t) = |\psi(t)\rangle\langle\psi(t)|$ relationship to calculate the negativity. By utilizing these relationships, we will obtain

$$\rho_{1,1}(t) = \frac{e^{-\frac{13}{4}i(t-t^*)}}{(|\alpha|^2+1)^2} \quad (31)$$

$$\rho_{1,2}(t) = \frac{\left(\frac{1}{4}-\frac{i}{4}\right)\alpha^* e^{-\frac{1}{4}i(2\sqrt{2}t^*+t^*+5t)}\left(\left(\sqrt{2}+(-1-i)\right)e^{i\sqrt{2}t^*}+(-1-i)-\sqrt{2}\right)}{(|\alpha|^2+1)^2} \quad (32)$$

$$\rho_{1,3}(t) = -\frac{\left(\frac{1}{4}+\frac{i}{4}\right)\alpha^* e^{-\frac{1}{4}i(2\sqrt{2}t^*+t^*+5t)}\left(\left(\sqrt{2}+(-1+i)\right)e^{i\sqrt{2}t^*}+(-1+i)-\sqrt{2}\right)}{(|\alpha|^2+1)^2} \quad (33)$$

$$\rho_{1,4}(t) = -\frac{(\alpha^*)^2 e^{-\frac{1}{4}i(3t^*+5t)}}{(|\alpha|^2+1)^2} \quad (34)$$

$$\rho_{2,1}(t) = \frac{\left(\frac{1}{4}+\frac{i}{4}\right)\alpha\left(-\left(\sqrt{2}+(1-i)\right)e^{i\sqrt{2}t}+(-1+i)+\sqrt{2}\right)e^{-\frac{1}{4}i\left((2\sqrt{2}-1)t-5t^*\right)}}{(|\alpha|^2+1)^2} \quad (35)$$

$$\rho_{2,2}(t) = \frac{1}{8(|\alpha|^2+1)^2} \alpha\left(\left(\sqrt{2}+(1-i)\right)e^{i\sqrt{2}t}+(1-i)-\sqrt{2}\right)\alpha^* e^{-\frac{1}{4}i(2\sqrt{2}-1)(t-t^*)}\left(\left(\sqrt{2}+(1+i)\right)e^{-i\sqrt{2}t^*}+(1+i)-\sqrt{2}\right) \quad (36)$$

$$\rho_{2,3}(t) = -\frac{1}{8(|\alpha|^2+1)^2} i\alpha\left(\left(\sqrt{2}+(1-i)\right)e^{i\sqrt{2}t}+(1-i)-\sqrt{2}\right)\alpha^* e^{-\frac{1}{4}i(2\sqrt{2}-1)(t-t^*)}\left(\left(\sqrt{2}+(1-i)\right)e^{-i\sqrt{2}t^*}+(1-i)-\sqrt{2}\right) \quad (37)$$

$$\rho_{2,4}(t) = \frac{\left(\frac{1}{4}+\frac{i}{4}\right)\alpha\left(\left(\sqrt{2}+(1-i)\right)e^{i\sqrt{2}t}+(1-i)-\sqrt{2}\right)(\alpha^*)^2 e^{-\frac{1}{4}i\left(3t^*+(2\sqrt{2}-1)t\right)}}{(|\alpha|^2+1)^2} \quad (38)$$

$$\rho_{3,1}(t) = \frac{\left(\frac{1}{4}-\frac{i}{4}\right)\alpha\left(\left(\sqrt{2}+(1+i)\right)e^{i\sqrt{2}t}+(1+i)-\sqrt{2}\right)e^{-\frac{1}{4}i\left((2\sqrt{2}-1)t-5t^*\right)}}{(|\alpha|^2+1)^2} \quad (39)$$

$$\rho_{3,2}(t)=\frac{1}{8(|\alpha|^2+1)^2} i\alpha\left(\left(\sqrt{2}+(1+i)\right)e^{i\sqrt{2}t}+(1+i)-\sqrt{2}\right)\alpha^* e^{-\frac{1}{4}i(2\sqrt{2}-1)(t-t^*)}\left(\left(\sqrt{2}+(1+i)\right)e^{-i\sqrt{2}t^*}+(1+i)-\sqrt{2}\right) \quad (40)$$

$$\rho_{3,3}(t)=\frac{1}{8(|\alpha|^2+1)^2} \alpha\left(\left(\sqrt{2}+(1+i)\right)e^{i\sqrt{2}t}+(1+i)-\sqrt{2}\right)\alpha^* e^{-\frac{1}{4}i(2\sqrt{2}-1)(t-t^*)}\left(\left(\sqrt{2}+(1-i)\right)e^{-i\sqrt{2}t^*}+(1-i)-\sqrt{2}\right) \quad (41)$$

$$\rho_{3,4}(t)=-\frac{1}{(|\alpha|^2+1)^2}\left(\frac{1}{4}-\frac{i}{4}\right)\alpha\left(\left(\sqrt{2}+(1+i)\right)e^{i\sqrt{2}t}+(1+i)-\sqrt{2}\right)(\alpha^*)^2 e^{-\frac{1}{4}i(3t^*+(2\sqrt{2}-1)t)} \quad (42)$$

$$\rho_{4,1}(t)=-\frac{\alpha^2 e^{\frac{1}{4}i(5t^*+3t)}}{(|\alpha|^2+1)^2} \quad (43)$$

$$\rho_{4,2}(t)=\frac{\left(\frac{1}{4}-\frac{i}{4}\right)\alpha^2\alpha^* e^{\frac{1}{4}i\left((2\sqrt{2}-1)t^*+3t\right)}\left(\left(\sqrt{2}+(1+i)\right)e^{-i\sqrt{2}t^*}+(1+i)-\sqrt{2}\right)}{(|\alpha|^2+1)^2} \quad (44)$$

$$\rho_{4,3}(t)=\frac{\left(\frac{1}{4}+\frac{i}{4}\right)\alpha^2\alpha^* e^{\frac{1}{4}i\left(3t-(1+2\sqrt{2})t^*\right)}\left(\left(\sqrt{2}+(-1+i)\right)e^{i\sqrt{2}t^*}+(-1+i)-\sqrt{2}\right)}{(|\alpha|^2+1)^2} \quad (45)$$

$$\rho_{4,4}(t)=\frac{\alpha^2(\alpha^*)^2 e^{\frac{3}{4}i(t-t^*)}}{(|\alpha|^2+1)^2} \quad (46)$$

In Fig. 2, we depict the negativity plot for subsystem $a$ in an isotropic $XXX$ state. The external fields on qubit $a$ are smaller than those on qubit $b$, for three $\alpha$ values. Similar to Fig. 1, fluctuations in negativity, reflecting entanglement variations, are evident. Entanglement death is absent. For $\alpha = 1$, the time-averaged entanglement is highest, followed by $\alpha = 2$, and $\alpha = 3$ has the lowest time average. Due to the smaller external fields on qubit $a$ compared to qubit $b$, the peak oscillations are consistent across all $\alpha$ values. These findings are derived from the following relationships

$$U(t)=\begin{pmatrix} e^{-\frac{5it}{4}} & 0 & 0 & 0 \\ 0 & \frac{1}{2}e^{\frac{it}{4}-\frac{it}{\sqrt{2}}}+\frac{1}{2}e^{\frac{it}{\sqrt{2}}+\frac{it}{4}} & \frac{\left(\frac{1}{2}+\frac{i}{2}\right)e^{\frac{it}{4}-\frac{it}{\sqrt{2}}}}{\sqrt{2}}-\frac{\left(\frac{1}{2}+\frac{i}{2}\right)e^{\frac{it}{\sqrt{2}}+\frac{it}{4}}}{\sqrt{2}} & 0 \\ 0 & \frac{\left(\frac{1}{2}-\frac{i}{2}\right)e^{\frac{it}{4}-\frac{it}{\sqrt{2}}}}{\sqrt{2}}-\frac{\left(\frac{1}{2}-\frac{i}{2}\right)e^{\frac{it}{\sqrt{2}}+\frac{it}{4}}}{\sqrt{2}} & \frac{1}{2}e^{\frac{it}{4}-\frac{it}{\sqrt{2}}}+\frac{1}{2}e^{\frac{it}{\sqrt{2}}+\frac{it}{4}} & 0 \\ 0 & 0 & 0 & e^{\frac{3it}{4}} \end{pmatrix} \quad (47)$$

and

$$|\psi(t)\rangle = U(t)|\psi(0)\rangle = -\frac{\alpha^2 e^{\frac{3it}{4}}|11\rangle}{|\alpha|^2+1}+\frac{e^{-\frac{5it}{4}}|00\rangle}{|\alpha|^2+1}+|01\rangle\left(\frac{\left(\frac{1}{2}+\frac{i}{2}\right)\alpha e^{\frac{it}{4}-\frac{it}{\sqrt{2}}}}{\sqrt{2}(|\alpha|^2+1)}-\frac{\alpha e^{\frac{it}{4}-\frac{it}{\sqrt{2}}}}{2(|\alpha|^2+1)}-\frac{\left(\frac{1}{2}+\frac{i}{2}\right)\alpha e^{\frac{it}{\sqrt{2}}+\frac{it}{4}}}{\sqrt{2}(|\alpha|^2+1)}-\frac{\alpha e^{\frac{it}{\sqrt{2}}+\frac{it}{4}}}{2(|\alpha|^2+1)}\right)+|10\rangle\left(-\frac{\left(\frac{1}{2}-\frac{i}{2}\right)\alpha e^{\frac{it}{4}-\frac{it}{\sqrt{2}}}}{\sqrt{2}(|\alpha|^2+1)}+\frac{\alpha e^{\frac{it}{4}-\frac{it}{\sqrt{2}}}}{2(|\alpha|^2+1)}+\frac{\left(\frac{1}{2}-\frac{i}{2}\right)\alpha e^{\frac{it}{\sqrt{2}}+\frac{it}{4}}}{\sqrt{2}(|\alpha|^2+1)}+\frac{\alpha e^{\frac{it}{\sqrt{2}}+\frac{it}{4}}}{2(|\alpha|^2+1)}\right) \quad (48)$$

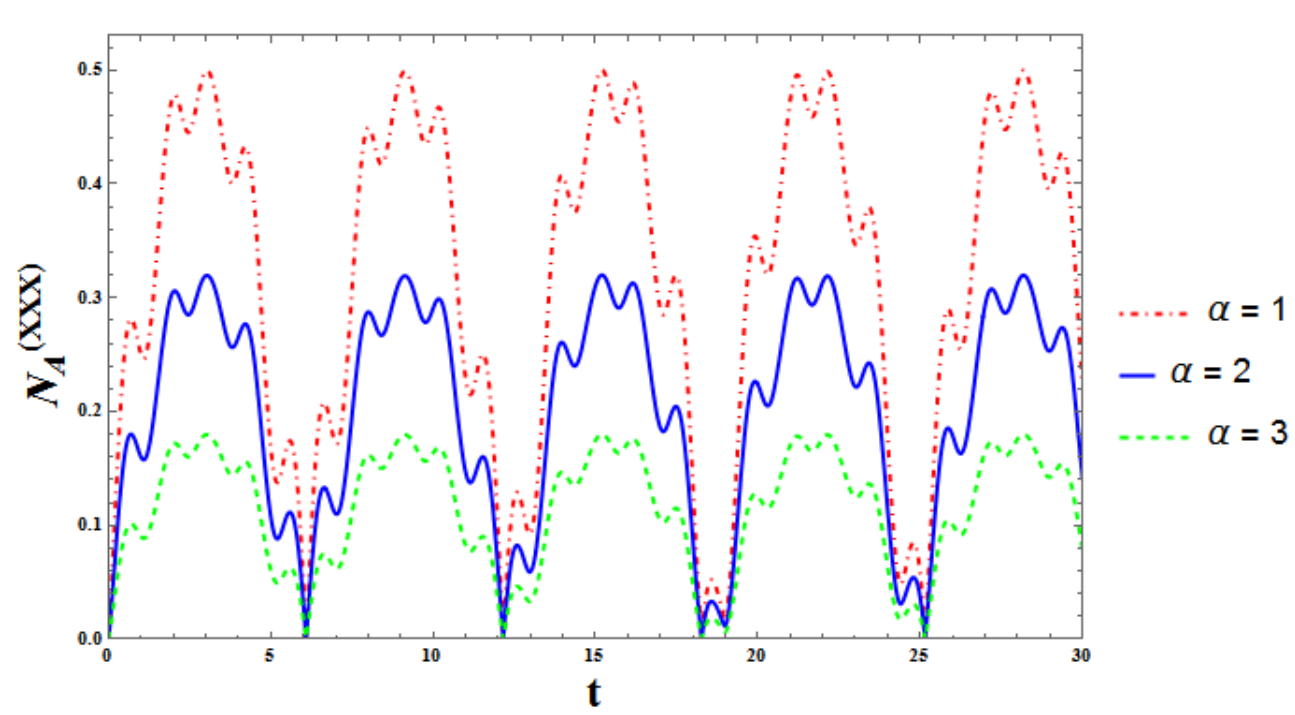

Figure 2: Time negativity plot for isotropic state $XXX$ ($J_x = J_y = J_z$) of subsystem (qubit) $a$ with three different $\alpha$ values and initial values of $J_x = 1, J_y = 1, J_z = 1, B_{z,a} = 1, B_{z,b} = 5, D_x = 0, D_y = 0, D_z = 1, \mathcal{N} = 2, \varphi = 0, \theta = \pi/4$. The magnetic fields applied to both qubits are in the z-axis direction, but they have different ($B_{z,a} < B_{z,b}$) magnitudes.

and in the following, we will use the $\rho(t) = |\psi(t)\rangle\langle\psi(t)|$ relationship to calculate the negativity. By utilizing these relationships, we will obtain

$$\rho_{1,1}(t) = \frac{e^{-\frac{5}{4}i(t-t^*)}}{(|\alpha|^2+1)^2} \quad (49)$$

$$\rho_{1,2}(t) = \frac{\alpha^* e^{-\frac{1}{4}i(6\sqrt{2}t^*+t^*+13t)}\left((-6+(5-i)\sqrt{2})e^{3i\sqrt{2}t^*}-6-(5-i)\sqrt{2}\right)}{12(|\alpha|^2+1)^2} \quad (50)$$

$$\rho_{1,3}(t) = \frac{\alpha^* e^{-\frac{1}{4}i(6\sqrt{2}t^*+t^*+13t)}\left((6+(3-i)\sqrt{2})e^{3i\sqrt{2}t^*}+6-(3-i)\sqrt{2}\right)}{12(|\alpha|^2+1)^2} \quad (51)$$

$$\rho_{1,4}(t) = -\frac{(\alpha^*)^2 e^{-\frac{1}{4}i(11t^*+13t)}}{(|\alpha|^2+1)^2} \quad (52)$$

$$\rho_{2,1}(t) = \frac{\alpha\left((-6-(5+i)\sqrt{2})e^{3i\sqrt{2}t}-6+(5+i)\sqrt{2}\right)e^{-\frac{1}{4}i\left((6\sqrt{2}-1)t-13t^*\right)}}{12(|\alpha|^2+1)^2} \quad (53)$$

$$\rho_{2,2}(t) = \frac{11\alpha\alpha^*}{9(|\alpha|-i)^2(|\alpha|+i)^2} - \frac{i(3\sqrt{2}-4i)\alpha e^{-3i\sqrt{2}t}\alpha^*}{36(|\alpha|-i)^2(|\alpha|+i)^2} + \frac{i(3\sqrt{2}+4i)\alpha e^{3i\sqrt{2}t}\alpha^*}{36(|\alpha|-i)^2(|\alpha|+i)^2} \quad (54)$$

$$\rho_{2,3}(t) = -\frac{\left(\frac{1}{18}+\frac{i}{18}\right)\alpha\alpha^*}{(|\alpha|-i)^2(|\alpha|+i)^2} + \frac{\left(\frac{1}{36}-\frac{i}{36}\right)\left((-9-8i)+(6+6i)\sqrt{2}\right)\alpha e^{-3i\sqrt{2}t}\alpha^*}{(|\alpha|-i)^2(|\alpha|+i)^2} - \frac{\left(\frac{1}{36}-\frac{i}{36}\right)\left((9+8i)+(6+6i)\sqrt{2}\right)\alpha e^{3i\sqrt{2}t}\alpha^*}{(|\alpha|-i)^2(|\alpha|+i)^2} \quad (55)$$

$$\rho_{2,4}(t) = \frac{1}{(|\alpha|^2+1)^2} \left(\frac{1}{12}-\frac{i}{12}\right)\alpha\left(\left((3+3i)+(2+3i)\sqrt{2}\right)e^{3i\sqrt{2}t} + (3+3i)-(2+3i)\sqrt{2}\right)(\alpha^*)^2 e^{-\frac{1}{4}i\left(11t^*+(6\sqrt{2}-1)t\right)} \quad (56)$$

$$\rho_{3,1}(t) = \frac{\alpha\left((6-(3+i)\sqrt{2})e^{3i\sqrt{2}t}+6+(3+i)\sqrt{2}\right)e^{-\frac{1}{4}i\left((6\sqrt{2}-1)t-13t^*\right)}}{12(|\alpha|^2+1)^2} \quad (57)$$

$$\rho_{3,2}(t) = -\frac{\left(\frac{1}{18}-\frac{i}{18}\right)\alpha\alpha^*}{(|\alpha|-i)^2(|\alpha|+i)^2} - \frac{\left(\frac{1}{36}-\frac{i}{36}\right)\left((8+9i)+(6+6i)\sqrt{2}\right)\alpha e^{-3i\sqrt{2}t}\alpha^*}{(|\alpha|-i)^2(|\alpha|+i)^2} + \frac{\left(\frac{1}{36}-\frac{i}{36}\right)\left((-8-9i)+(6+6i)\sqrt{2}\right)\alpha e^{3i\sqrt{2}t}\alpha^*}{(|\alpha|-i)^2(|\alpha|+i)^2} \quad (58)$$

$$\rho_{3,3}(t) = \frac{7\alpha\alpha^*}{9(|\alpha|-i)^2(|\alpha|+i)^2} + \frac{(4+3i\sqrt{2})\alpha e^{-3i\sqrt{2}t}\alpha^*}{36(|\alpha|-i)^2(|\alpha|+i)^2} + \frac{(4-3i\sqrt{2})\alpha e^{3i\sqrt{2}t}\alpha^*}{36(|\alpha|-i)^2(|\alpha|+i)^2} \quad (59)$$

$$\rho_{3,4}(t) = \frac{1}{(|\alpha|^2+1)^2} \left(\frac{1}{12}-\frac{i}{12}\right)\alpha\left(\left((-3-3i)+(1+2i)\sqrt{2}\right)e^{3i\sqrt{2}t} + (-3-3i)-(1+2i)\sqrt{2}\right)(\alpha^*)^2 e^{-\frac{1}{4}i\left(11t^*+(6\sqrt{2}-1)t\right)} \quad (60)$$

$$\rho_{4,1}(t) = -\frac{\alpha^2 e^{\frac{1}{4}i(13t^*+11t)}}{(|\alpha|^2+1)^2} \quad (61)$$

$$\rho_{4,2}(t) = \frac{1}{(|\alpha|^2+1)^2}\left(\frac{1}{12}+\frac{i}{12}\right) \alpha^2\alpha^* e^{\frac{1}{4}i\left((6\sqrt{2}-1)t^*+11t\right)}\left(\left((3-3i)+(2-3i)\sqrt{2}\right)e^{-3i\sqrt{2}t^*} + (3-3i)-(2-3i)\sqrt{2}\right) \quad (62)$$

$$\rho_{4,3}(t) = \frac{-1}{(|\alpha|^2+1)^2} \left(\frac{1}{12}-\frac{i}{12}\right)\alpha^2\alpha^* e^{\frac{1}{4}i\left(11t-(1+6\sqrt{2})t^*\right)}\left(\left((3+3i)+(2+i)\sqrt{2}\right)e^{3i\sqrt{2}t^*} + (3+3i)-(2+i)\sqrt{2}\right) \quad (63)$$

$$\rho_{4,4}(t) = \frac{\alpha^2(\alpha^*)^2 e^{\frac{11}{4}i(t-t^*)}}{(|\alpha|^2+1)^2} \quad (64)$$

In Fig. 3, the negativity plot for subsystem $a$ in the $XXX$ isotropic state is shown with external fields on qubit $a$ exceeding those on qubit $b$, for three $\alpha$ values. Similar to Fig. 2, fluctuations in negativity, representing entanglement changes, are observable. Just as in the previous case, entanglement doesn't vanish. For $\alpha = 1$, the average entanglement is highest, followed by $\alpha = 2$, with $\alpha = 3$ having the lowest average. Because the external fields on qubit $a$ are greater than those on qubit $b$, the peak oscillations are consistent across all three $\alpha$ values. The oscillatory behavior of these peaks in Fig. 3 mirrors that of Fig. 2. These peaks represent the relative maxima of negativity that occur within a specific repeating interval, highlighting the points of highest negativity in the cycle. The relative maxima arise due to the fact that the magnetic fields that are applied to the two qubits are not identical and instead vary from one another. This difference in the magnetic fields is what leads to the observed relative maxima in the system. These insights are based on the given relationships

$$U_{1,2}(t) = U_{1,3}(t) = U_{1,4}(t) = U_{2,1}(t) = U_{2,4}(t) = U_{3,1}(t) = U_{3,4}(t) = U_{4,1}(t) = U_{4,2}(t) = U_{4,3}(t) = 0 \quad (65)$$

$$U_{1,1}(t) = e^{-\frac{13it}{4}} \quad (66)$$

$$U_{2,2}(t) = \frac{1}{2}e^{\frac{it}{4}-\frac{3it}{\sqrt{2}}} + \frac{1}{3}\sqrt{2}e^{\frac{it}{4}-\frac{3it}{\sqrt{2}}} + \frac{1}{2}e^{\frac{3it}{\sqrt{2}}+\frac{it}{4}} - \frac{1}{3}\sqrt{2}e^{\frac{3it}{\sqrt{2}}+\frac{it}{4}} \quad (67)$$

$$U_{2,3}(t) = \frac{\left(\frac{1}{6}+\frac{i}{6}\right)e^{\frac{it}{4}-\frac{3it}{\sqrt{2}}}}{\sqrt{2}} - \frac{\left(\frac{1}{6}+\frac{i}{6}\right)e^{\frac{3it}{\sqrt{2}}+\frac{it}{4}}}{\sqrt{2}} \quad (68)$$

$$U_{3,2}(t) = \frac{\left(\frac{1}{6}-\frac{i}{6}\right)e^{\frac{it}{4}-\frac{3it}{\sqrt{2}}}}{\sqrt{2}} - \frac{\left(\frac{1}{6}-\frac{i}{6}\right)e^{\frac{3it}{\sqrt{2}}+\frac{it}{4}}}{\sqrt{2}} \quad (69)$$

$$U_{3,3}(t) = \frac{1}{2}e^{\frac{it}{4}-\frac{3it}{\sqrt{2}}} - \frac{1}{3}\sqrt{2}e^{\frac{it}{4}-\frac{3it}{\sqrt{2}}} + \frac{1}{2}e^{\frac{3it}{\sqrt{2}}+\frac{it}{4}} + \frac{1}{3}\sqrt{2}e^{\frac{3it}{\sqrt{2}}+\frac{it}{4}} \quad (70)$$

$$U_{4,4}(t) = e^{\frac{11it}{4}} \quad (71)$$

and

$$|\psi(t)\rangle = U(t)|\psi(0)\rangle = -\frac{\alpha^2 e^{\frac{11it}{4}}|11\rangle}{|\alpha|^2+1} + \frac{e^{-\frac{13it}{4}}|00\rangle}{|\alpha|^2+1} + |01\rangle \left( \frac{\left(\frac{5}{6}+\frac{i}{6}\right)\alpha e^{\frac{it}{4}-\frac{3it}{\sqrt{2}}}}{\sqrt{2}(|\alpha|^2+1)} - \frac{\alpha e^{\frac{it}{4}-\frac{3it}{\sqrt{2}}}}{2(|\alpha|^2+1)} - \frac{\left(\frac{5}{6}+\frac{i}{6}\right)\alpha e^{\frac{3it}{\sqrt{2}}+\frac{it}{4}}}{\sqrt{2}(|\alpha|^2+1)} - \frac{\alpha e^{\frac{3it}{\sqrt{2}}+\frac{it}{4}}}{2(|\alpha|^2+1)} \right) + |10\rangle \left( \frac{\left(\frac{1}{2}+\frac{i}{6}\right)\alpha e^{\frac{it}{4}-\frac{3it}{\sqrt{2}}}}{\sqrt{2}(|\alpha|^2+1)} + \frac{\alpha e^{\frac{it}{4}-\frac{3it}{\sqrt{2}}}}{2(|\alpha|^2+1)} - \frac{\left(\frac{1}{2}+\frac{i}{6}\right)\alpha e^{\frac{3it}{\sqrt{2}}+\frac{it}{4}}}{\sqrt{2}(|\alpha|^2+1)} + \frac{\alpha e^{\frac{3it}{\sqrt{2}}+\frac{it}{4}}}{2(|\alpha|^2+1)} \right) \quad (72)$$

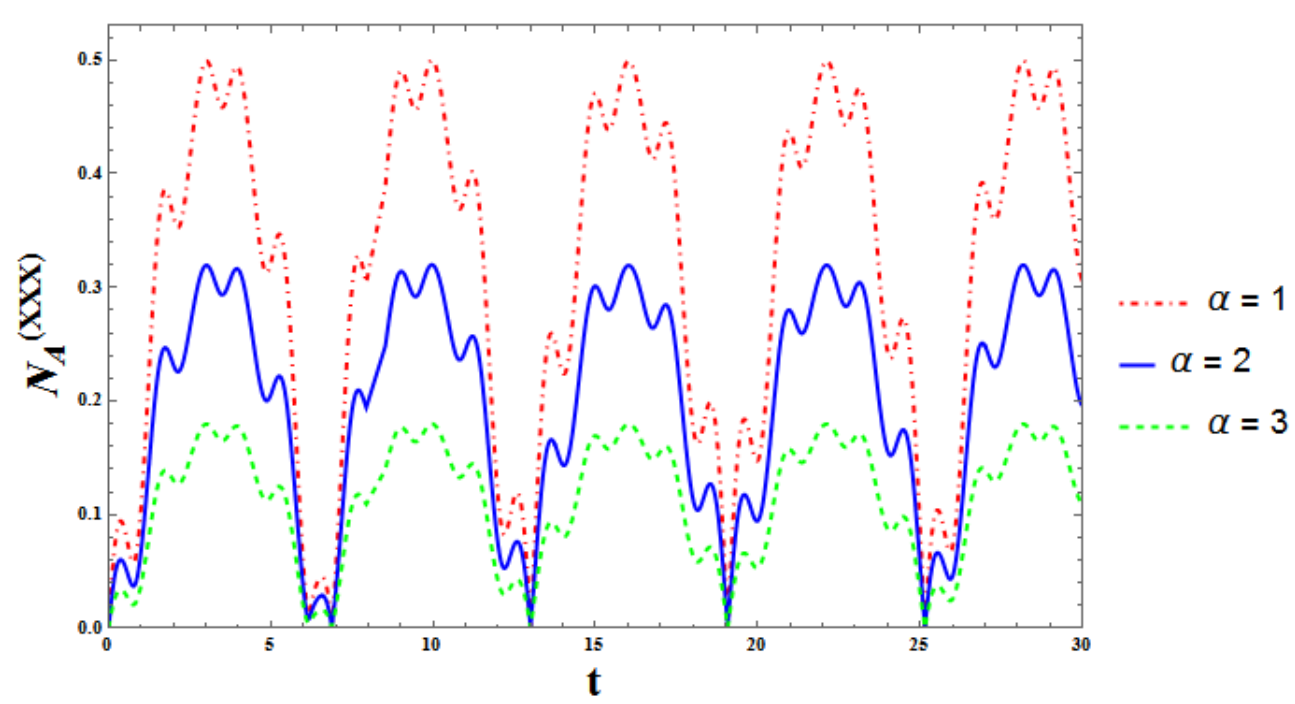

Figure 3: Time negativity plot for subsystem (qubit) $a$ in the $XXX$ ($J_x = J_y = J_z$) isotropic state for three $\alpha$ values with initial values of $J_x = 1, J_y = 1, J_z = 1, B_{z,a} = 5, B_{z,b} = 1, D_x = 0, D_y = 0, D_z = 1, \mathcal{N} = 2, \varphi = 0, \theta = \pi/4$. The magnetic fields applied to both qubits are in the z-axis direction, but they have different ($B_{z,a} > B_{z,b}$) magnitudes.

and in the following, we will use the $\rho(t) = |\psi(t)\rangle\langle\psi(t)|$ relationship to calculate the negativity.

By using these relationships, we will obtain

$$\rho_{1,1}(t) = \frac{e^{\frac{13it^*}{4}-\frac{13it}{4}}}{(|\alpha|^2+1)^2} \quad (73)$$

$$\rho_{1,2}(t) = \frac{\left(\frac{1}{12}-\frac{i}{12}\right)\left((-3-3i)+(1+2i)\sqrt{2}\right)\alpha^* e^{-\frac{3it^*}{\sqrt{2}}-\frac{it^*}{4}-\frac{13it}{4}}}{(|\alpha|-i)^2(|\alpha|+i)^2} - \frac{\left(\frac{1}{12}-\frac{i}{12}\right)\left((3+3i)+(1+2i)\sqrt{2}\right)\alpha^* e^{\frac{3it^*}{\sqrt{2}}-\frac{it^*}{4}-\frac{13it}{4}}}{(|\alpha|-i)^2(|\alpha|+i)^2} \quad (74)$$

$$\rho_{1,3}(t) = \frac{\left(\frac{1}{12}-\frac{i}{12}\right)\left((3+3i)+(2+3i)\sqrt{2}\right)\alpha^* e^{-\frac{3it^*}{\sqrt{2}}-\frac{it^*}{4}-\frac{13it}{4}}}{(|\alpha|-i)^2(|\alpha|+i)^2} - \frac{\left(\frac{1}{12}-\frac{i}{12}\right)\left((-3-3i)+(2+3i)\sqrt{2}\right)\alpha^* e^{\frac{3it^*}{\sqrt{2}}-\frac{it^*}{4}-\frac{13it}{4}}}{(|\alpha|-i)^2(|\alpha|+i)^2} \quad (75)$$

$$\rho_{1,4}(t) = -\frac{(\alpha^*)^2 e^{-\frac{11it^*}{4}-\frac{13it}{4}}}{(|\alpha|^2+1)^2} \quad (76)$$

$$\rho_{2,1}(t) = \frac{1}{(|\alpha|-i)^2(|\alpha|+i)^2}\left(\frac{1}{12}-\frac{i}{12}\right)\alpha\left(-(3+3i)e^{3i\sqrt{2}t}+(2+i)\sqrt{2}e^{3i\sqrt{2}t}+(-3-3i)-(2+i)\sqrt{2}\right)e^{\frac{13it^*}{4}-\frac{1}{4}i(6\sqrt{2}-1)t} \quad (77)$$

$$\rho_{2,2}(t) = \frac{7\alpha\alpha^*}{9(|\alpha|-i)^2(|\alpha|+i)^2} + \frac{(4-3i\sqrt{2})\alpha e^{-3i\sqrt{2}t}\alpha^*}{36(|\alpha|-i)^2(|\alpha|+i)^2} + \frac{(4+3i\sqrt{2})\alpha e^{3i\sqrt{2}t}\alpha^*}{36(|\alpha|-i)^2(|\alpha|+i)^2} \quad (78)$$

$$\rho_{2,3}(t) = -\frac{\left(\frac{1}{18}+\frac{i}{18}\right)\alpha\alpha^*}{(|\alpha|-i)^2(|\alpha|+i)^2} - \frac{\left(\frac{1}{36}-\frac{i}{36}\right)\left((9+8i)+(6+6i)\sqrt{2}\right)\alpha e^{-3i\sqrt{2}t}\alpha^*}{(|\alpha|-i)^2(|\alpha|+i)^2} + \frac{\left(\frac{1}{36}-\frac{i}{36}\right)\left((-9-8i)+(6+6i)\sqrt{2}\right)\alpha e^{3i\sqrt{2}t}\alpha^*}{(|\alpha|-i)^2(|\alpha|+i)^2} \quad (79)$$

$$\rho_{2,4}(t) = \frac{-1}{(|\alpha|-i)^2(|\alpha|+i)^2}\left(\frac{1}{12}-\frac{i}{12}\right)\alpha e^{-\frac{1}{4}i(6\sqrt{2}-1)t-\frac{11it}{4}}\left(-(3+3i)e^{3i\sqrt{2}t}+(2+i)\sqrt{2}e^{3i\sqrt{2}t}+(-3-3i)-(2+i)\sqrt{2}\right)(\alpha^*)^2 \quad (80)$$

$$\rho_{3,1}(t) = \frac{1}{(|\alpha|-i)^2(|\alpha|+i)^2}\left(\frac{1}{12}-\frac{i}{12}\right)\alpha e^{\frac{13it}{4}-\frac{1}{4}i(6\sqrt{2}-1)t}\left((3+3i)e^{3i\sqrt{2}t}+(3+2i)\sqrt{2}e^{3i\sqrt{2}t}+(3+3i)-(3+2i)\sqrt{2}\right) \quad (81)$$

$$\rho_{3,2}(t) = -\frac{\left(\frac{1}{18}-\frac{i}{18}\right)\alpha\alpha^*}{(|\alpha|-i)^2(|\alpha|+i)^2} + \frac{\left(\frac{1}{36}-\frac{i}{36}\right)\left((-8-9i)+(6+6i)\sqrt{2}\right)\alpha e^{-3i\sqrt{2}t}\alpha^*}{(|\alpha|-i)^2(|\alpha|+i)^2} - \frac{\left(\frac{1}{36}-\frac{i}{36}\right)\left((8+9i)+(6+6i)\sqrt{2}\right)\alpha e^{3i\sqrt{2}t}\alpha^*}{(|\alpha|-i)^2(|\alpha|+i)^2} \quad (82)$$

$$\rho_{3,3}(t) = \frac{11\alpha\alpha^*}{9(|\alpha|-i)^2(|\alpha|+i)^2} + \frac{i(3\sqrt{2}+4i)\alpha e^{-3i\sqrt{2}t}\alpha^*}{36(|\alpha|-i)^2(|\alpha|+i)^2} - \frac{i(3\sqrt{2}-4i)\alpha e^{3i\sqrt{2}t}\alpha^*}{36(|\alpha|-i)^2(|\alpha|+i)^2} \quad (83)$$

$$\rho_{3,4}(t) = \frac{-1}{(|\alpha|-i)^2(|\alpha|+i)^2}\left(\frac{1}{12}-\frac{i}{12}\right)\alpha e^{-\frac{1}{4}i(6\sqrt{2}-1)t-\frac{11it}{4}}\left((3+3i)e^{3i\sqrt{2}t}+(3+2i)\sqrt{2}e^{3i\sqrt{2}t}+(3+3i)-(3+2i)\sqrt{2}\right)(\alpha^*)^2 \quad (84)$$

$$\rho_{4,1}(t) = -\frac{\alpha^2 e^{6it}}{(|\alpha|^2+1)^2} \quad (85)$$

$$\rho_{4,2}(t) = \frac{1}{(|\alpha|-i)^2(|\alpha|+i)^2}\left(\frac{1}{12}-\frac{i}{12}\right)\alpha^2 e^{\frac{11it}{4}-\frac{1}{4}i(1+6\sqrt{2})t}\left((3+3i)e^{3i\sqrt{2}t}+(1+2i)\sqrt{2}e^{3i\sqrt{2}t}+(3+3i)-(1+2i)\sqrt{2}\right)\alpha^* \quad (86)$$

$$\rho_{4,3}(t) = \frac{1}{(|\alpha|-i)^2(|\alpha|+i)^2}\left(\frac{1}{12}-\frac{i}{12}\right)\alpha^2 e^{\frac{1}{4}i(6\sqrt{2}-1)t-3i\sqrt{2}t+\frac{11it}{4}}\left(-(3+3i)e^{3i\sqrt{2}t}+(2+3i)\sqrt{2}e^{3i\sqrt{2}t}+(-3-3i)-(2+3i)\sqrt{2}\right)\alpha^* \quad (87)$$

$$\rho_{4,4}(t) = \frac{\alpha^2(\alpha^*)^2}{(|\alpha|^2+1)^2} \quad (88)$$

In Fig. 4, we illustrate the negativity plot for subsystem $\alpha$ in the $XYZ$ anisotropic state across three $\alpha$ values. In this graph, fluctuations in the negativity scale (representing entanglement) appear irregular. The maximum peaks are still associated with $\alpha = 1$, followed by $\alpha = 2$, and the minimum peaks align with $\alpha = 3$. The time-averaged entanglement is highest for $\alpha = 1$ (0.30), next is $\alpha = 2$ (0.256), and lowest for $\alpha = 3$ (0.205). These conclusions stem from the following relationships

$$U_{1,2}(t)=U_{1,3}(t)=U_{2,1}(t)=U_{2,4}(t)=U_{3,1}(t)=U_{3,4}(t)=U_{4,2}(t)=U_{4,3}(t)=0 \quad (89)$$

$$U_{1,1}(t)=\frac{1}{2}e^{-\frac{1}{4}i\sqrt{17}t-\frac{3it}{4}}+\frac{2e^{-\frac{1}{4}i\sqrt{17}t-\frac{3it}{4}}}{\sqrt{17}}+\frac{1}{2}e^{\frac{1}{4}i\sqrt{17}t-\frac{3it}{4}}-\frac{2e^{\frac{1}{4}i\sqrt{17}t-\frac{3it}{4}}}{\sqrt{17}} \quad (90)$$

$$U_{1,4}(t)=\frac{e^{\frac{1}{4}i\sqrt{17}t-\frac{3it}{4}}}{2\sqrt{17}}-\frac{e^{-\frac{1}{4}i\sqrt{17}t-\frac{3it}{4}}}{2\sqrt{17}} \quad (91)$$

$$U_{2,2}(t)=\frac{1}{2}e^{\frac{3it}{4}-\frac{1}{4}i\sqrt{13}t}+\frac{1}{2}e^{\frac{1}{4}i\sqrt{13}t+\frac{3it}{4}} \quad (92)$$

$$U_{2,3}(t)=\frac{\left(\frac{3}{2}+i\right)e^{\frac{3it}{4}-\frac{1}{4}i\sqrt{13}t}}{\sqrt{13}}-\frac{\left(\frac{3}{2}+i\right)e^{\frac{1}{4}i\sqrt{13}t+\frac{3it}{4}}}{\sqrt{13}} \quad (93)$$

$$U_{3,2}(t)=\frac{\left(\frac{3}{2}-i\right)e^{\frac{3it}{4}-\frac{1}{4}i\sqrt{13}t}}{\sqrt{13}}-\frac{\left(\frac{3}{2}-i\right)e^{\frac{1}{4}i\sqrt{13}t+\frac{3it}{4}}}{\sqrt{13}} \quad (94)$$

$$U_{3,3}(t)=\frac{1}{2}e^{\frac{3it}{4}-\frac{1}{4}i\sqrt{13}t}+\frac{1}{2}e^{\frac{1}{4}i\sqrt{13}t+\frac{3it}{4}} \quad (95)$$

$$U_{4,1}(t)=\frac{e^{\frac{1}{4}i\sqrt{17}t-\frac{3it}{4}}}{2\sqrt{17}}-\frac{e^{-\frac{1}{4}i\sqrt{17}t-\frac{3it}{4}}}{2\sqrt{17}} \quad (96)$$

$$U_{4,4}(t)=\frac{1}{2}e^{-\frac{1}{4}i\sqrt{17}t-\frac{3it}{4}}-\frac{2e^{-\frac{1}{4}i\sqrt{17}t-\frac{3it}{4}}}{\sqrt{17}}+\frac{1}{2}e^{\frac{1}{4}i\sqrt{17}t-\frac{3it}{4}}+\frac{2e^{\frac{1}{4}i\sqrt{17}t-\frac{3it}{4}}}{\sqrt{17}} \quad (97)$$

and

$$|\psi(t)\rangle = U(t)|\psi(0)\rangle =$$

$$\left(\frac{e^{-\frac{1}{4}i\sqrt{17}t-\frac{3it}{4}}\alpha^2}{2\sqrt{17}(|\alpha|^2+1)}-\frac{e^{\frac{1}{4}i\sqrt{17}t-\frac{3it}{4}}\alpha^2}{2\sqrt{17}(|\alpha|^2+1)}+\frac{2e^{-\frac{1}{4}i\sqrt{17}t-\frac{3it}{4}}}{\sqrt{17}(|\alpha|^2+1)}+\frac{e^{-\frac{1}{4}i\sqrt{17}t-\frac{3it}{4}}}{2(|\alpha|^2+1)}-\frac{2e^{\frac{1}{4}i\sqrt{17}t-\frac{3it}{4}}}{\sqrt{17}(|\alpha|^2+1)}+\frac{e^{\frac{1}{4}i\sqrt{17}t-\frac{3it}{4}}}{2(|\alpha|^2+1)}\right)|00\rangle$$
$$+\left(\frac{\left(\frac{3}{2}+i\right)e^{\frac{3it}{4}-\frac{1}{4}i\sqrt{13}t}\alpha}{\sqrt{13}(|\alpha|^2+1)}-\frac{e^{\frac{3it}{4}-\frac{1}{4}i\sqrt{13}t}\alpha}{2(|\alpha|^2+1)}-\frac{\left(\frac{3}{2}+i\right)e^{\frac{1}{4}i\sqrt{13}t+\frac{3it}{4}}\alpha}{\sqrt{13}(|\alpha|^2+1)}-\frac{e^{\frac{1}{4}i\sqrt{13}t+\frac{3it}{4}}\alpha}{2(|\alpha|^2+1)}\right)|01\rangle$$
$$+\left(-\frac{\left(\frac{3}{2}-i\right)e^{\frac{3it}{4}-\frac{1}{4}i\sqrt{13}t}\alpha}{\sqrt{13}(|\alpha|^2+1)}+\frac{e^{\frac{3it}{4}-\frac{1}{4}i\sqrt{13}t}\alpha}{2(|\alpha|^2+1)}+\frac{\left(\frac{3}{2}-i\right)e^{\frac{1}{4}i\sqrt{13}t+\frac{3it}{4}}\alpha}{\sqrt{13}(|\alpha|^2+1)}+\frac{e^{\frac{1}{4}i\sqrt{13}t+\frac{3it}{4}}\alpha}{2(|\alpha|^2+1)}\right)|10\rangle+$$
$$\left(\frac{2e^{-\frac{1}{4}i\sqrt{17}t-\frac{3it}{4}}\alpha^2}{\sqrt{17}(|\alpha|^2+1)}-\frac{e^{-\frac{1}{4}i\sqrt{17}t-\frac{3it}{4}}\alpha^2}{2(|\alpha|^2+1)}-\frac{2e^{\frac{1}{4}i\sqrt{17}t-\frac{3it}{4}}\alpha^2}{\sqrt{17}(|\alpha|^2+1)}-\frac{e^{\frac{1}{4}i\sqrt{17}t-\frac{3it}{4}}\alpha^2}{2(|\alpha|^2+1)}-\frac{e^{-\frac{1}{4}i\sqrt{17}t-\frac{3it}{4}}}{2\sqrt{17}(|\alpha|^2+1)}+\frac{e^{\frac{1}{4}i\sqrt{17}t-\frac{3it}{4}}}{2\sqrt{17}(|\alpha|^2+1)}\right)|11\rangle$$
(98)

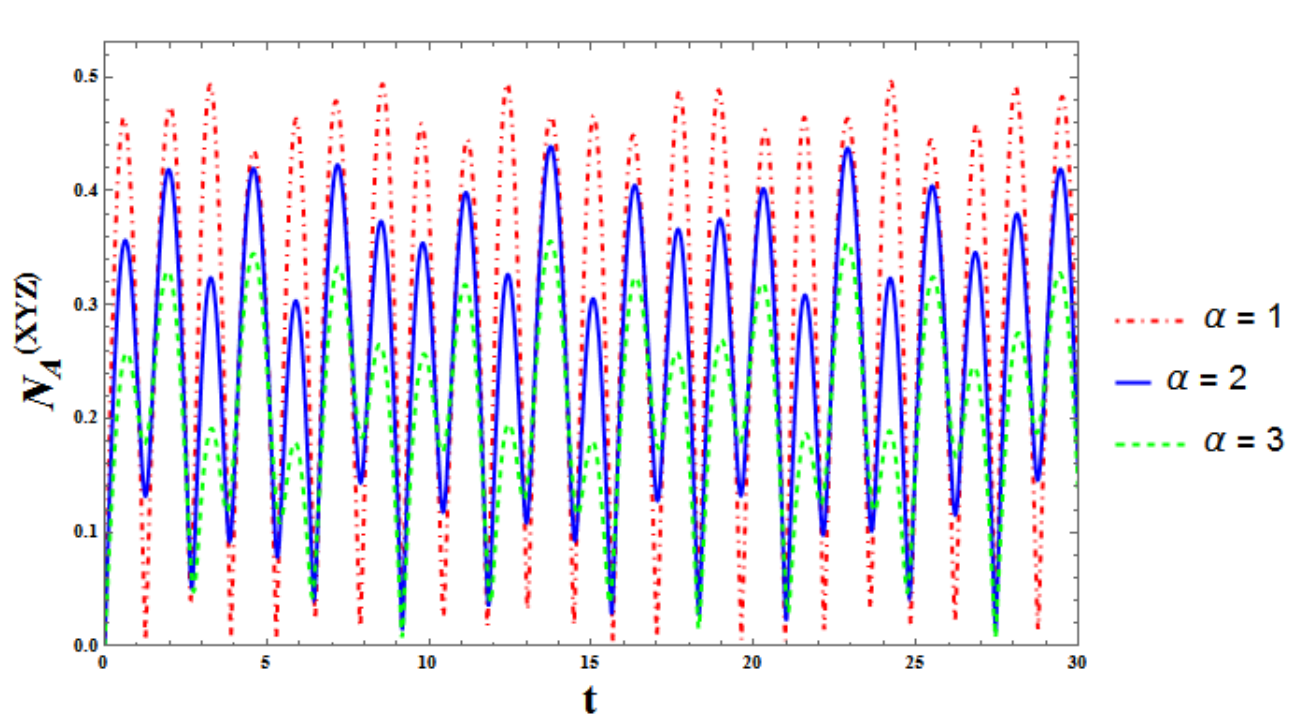

Figure 4: Time negativity diagram for $XYZ$ ($J_x \neq J_y \neq J_z$) anisotropic state of subsystem (qubit) $a$ for three different values of $\alpha$, with initial values of $J_x=1, J_y=2, J_z=3, B_{z,a}=1, B_{z,b}=1, D_x=0, D_y=0, D_z=1, \mathcal{N}=$

2, $\varphi = 0, \theta = \pi/4$. The magnetic fields applied to both qubits are equal ($B_{z,a} = B_{z,b}$) and oriented in the z-axis direction.

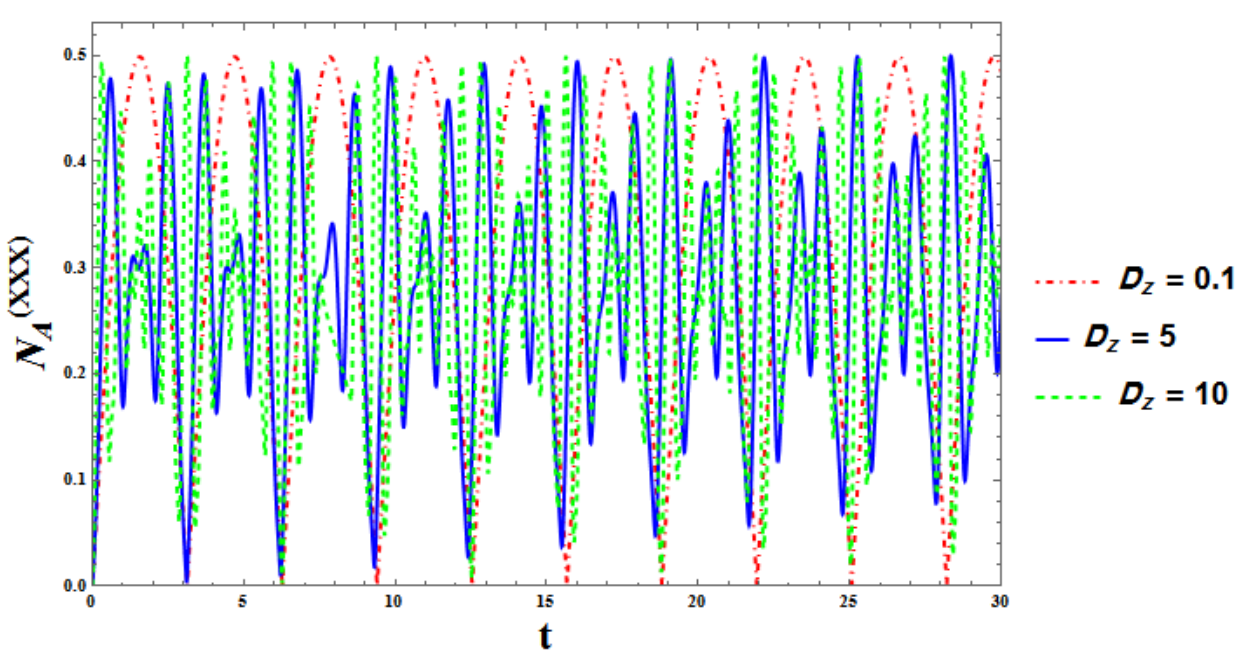

Figure 5: Time negativity diagram for isotropic state $XXX$ ($J_x = J_y = J_z$) of subsystem (qubit) $a$ for three different values of $D_z$ and initial values of $\alpha = 1, J_x = 1, J_y = 1, J_z = 1, B_{z,a} = 1, B_{z,b} = 1, D_x = 0, D_y = 0, \mathcal{N} = 2, \varphi = 0, \theta = \pi/4$. The magnetic fields applied to both qubits are equal ($B_{z,a} = B_{z,b}$) and oriented in the z-axis direction.

and in the following, we will use the $\rho(t) = |\psi(t)\rangle\langle\psi(t)|$ relationship to calculate the negativity. As the density matrix arrays are excessively large, we have refrained from documenting them.

In Fig. 5, we depicted the negativity plot of subsystem $a$ for the $XXX$ isotropic state across three different $D_z$ values. This graph reveals that as $D_z$ increases, the periodic fluctuations in negativity, or rather entanglement, intensify. However, the peak and trough values remain close for all $D_z$ values. The temporal evolution of negativity exhibits a more consistent pattern at lower $D_z$ values. The highest time-averaged entanglement occurs at $D_z = 0.1$ (0.319), followed by $D_z = 5$ (0.284), with the lowest at $D_z = 10$ (0.283). Consequently, the $DM$ interaction has raised the overall average entanglement. These findings are derived from the following relationships

$$U_{1,2}(t) = U_{1,3}(t) = U_{1,4}(t) = U_{2,1}(t) = U_{2,4}(t) = U_{3,1}(t) = U_{3,4}(t) = U_{4,1}(t) = U_{4,2}(t) = U_{4,3}(t) = 0 \quad (99)$$

$$U_{1,1}(t) = e^{-\frac{5it}{4}} \quad (100)$$

$$U_{2,2}(t) = \frac{1}{2} e^{\frac{1}{4}\left(-2\sqrt{-D_z^2-1}+i\right)t} + \frac{1}{2} e^{\frac{1}{4}\left(2\sqrt{-D_z^2-1}+i\right)t} \quad (101)$$

$$U_{2,3}(t) = \frac{(2D_z-2i)e^{\frac{1}{4}\left(2\sqrt{-D_z^2-1}+i\right)t}}{4\sqrt{-D_z^2-1}} - \frac{(2D_z-2i)e^{\frac{1}{4}\left(-2\sqrt{-D_z^2-1}+i\right)t}}{4\sqrt{-D_z^2-1}} \quad (102)$$

$$U_{3,2}(t) = \frac{(-2D_z-2i)e^{\frac{1}{4}\left(2\sqrt{-D_z^2-1}+i\right)t}}{4\sqrt{-D_z^2-1}} - \frac{(-2D_z-2i)e^{\frac{1}{4}\left(-2\sqrt{-D_z^2-1}+i\right)t}}{4\sqrt{-D_z^2-1}} \quad (103)$$

$$U_{3,3}(t) = \frac{1}{2} e^{\frac{1}{4}\left(-2\sqrt{-D_z^2-1}+i\right)t} + \frac{1}{2} e^{\frac{1}{4}\left(2\sqrt{-D_z^2-1}+i\right)t} \quad (104)$$

$$U_{4,4}(t) = e^{\frac{3it}{4}} \quad (105)$$

and

$$|\psi(t)\rangle = U(t)|\psi(0)\rangle = -\frac{D_z e^{\frac{1}{4}\left(-2\sqrt{-D_z^2-1}+i\right)t}|01\rangle}{4\sqrt{-D_z^2-1}} + \frac{i e^{\frac{1}{4}\left(-2\sqrt{-D_z^2-1}+i\right)t}|01\rangle}{4\sqrt{-D_z^2-1}} - \frac{1}{4} e^{\frac{1}{4}\left(-2\sqrt{-D_z^2-1}+i\right)t}|01\rangle + \frac{D_z e^{\frac{1}{4}\left(2\sqrt{-D_z^2-1}+i\right)t}|01\rangle}{4\sqrt{-D_z^2-1}} - \frac{i e^{\frac{1}{4}\left(2\sqrt{-D_z^2-1}+i\right)t}|01\rangle}{4\sqrt{-D_z^2-1}} - \frac{1}{4} e^{\frac{1}{4}\left(2\sqrt{-D_z^2-1}+i\right)t}|01\rangle - \frac{D_z e^{\frac{1}{4}\left(-2\sqrt{-D_z^2-1}+i\right)t}|10\rangle}{4\sqrt{-D_z^2-1}} -$$

$$\frac{ie^{\frac{1}{4}\left(-2\sqrt{-D_z^2-1}+i\right)t}|10\rangle}{4\sqrt{-D_z^2-1}} + \frac{1}{4}e^{\frac{1}{4}\left(-2\sqrt{-D_z^2-1}+i\right)t}|10\rangle + \frac{D_z e^{\frac{1}{4}\left(2\sqrt{-D_z^2-1}+i\right)t}|10\rangle}{4\sqrt{-D_z^2-1}} + \frac{ie^{\frac{1}{4}\left(2\sqrt{-D_z^2-1}+i\right)t}|10\rangle}{4\sqrt{-D_z^2-1}} + \frac{1}{4}e^{\frac{1}{4}\left(2\sqrt{-D_z^2-1}+i\right)t}|10\rangle + \frac{1}{2}e^{-\frac{5it}{4}}|00\rangle - \frac{1}{2}e^{\frac{3it}{4}}|11\rangle \quad (106)$$

and in the following, we will use the $\rho(t) = |\psi(t)\rangle\langle\psi(t)|$ relationship to calculate the negativity. As the density matrix arrays are excessively large, we have refrained from documenting them. The conclusions drawn from the present study exhibit a strong correlation with the research findings of Chamgordani et al [18] as well as Zhang et al [19], indicating a consistent and harmonious relationship between the respective research outcomes.

Quantum entanglement dynamics in qubit-qubit systems is a key area in quantum mechanics and information science. Here are some applications:

- Quantum Computing: Entangled qubits are essential for quantum computers, enabling faster algorithms like Shor's for factoring and Grover's for searching.
- Quantum Cryptography: Entanglement is vital for protocols like BB84 and E91, allowing secure key distribution and detection of interceptions.
- Quantum Teleportation: It enables the transfer of quantum states between distant qubits, facilitating communication without direct transmission.
- Quantum State Transfer: Entangled qubits facilitate long-distance quantum state transfers in quantum communication networks.
- Quantum Sensors: Entangled systems improve measurement precision in sensors, reducing noise in applications like gravitational wave detectors.
- Quantum Metrology: They enhance parameter estimation, crucial for measurements in astronomy and navigation.
- Quantum Simulation: These systems can simulate complex quantum phenomena, aiding studies in condensed matter physics.
- Testing Quantum Principles: Experiments with entangled qubits test fundamental quantum mechanics concepts like Bell's theorem.
- Quantum Error Correction: They are crucial for creating error-correcting codes that protect quantum information from noise.
- Quantum Games: Entanglement influences game theory strategies and decision-making in competitive scenarios.

## 5 Conclusion

This article delves into the analysis of the quantum coherence dynamics within a complex qubit-qubit system in the $XXX$ isotropic Heisenberg model and the anisotropic $XYZ$ model, considering the influence of $DM$ interaction and independent external magnetic fields acting on both qubits. The initial state is a spin coherent state, with negativity utilized as the metric for measuring entanglement. The investigation focuses on understanding how the entanglement dynamics are affected by the interplay of $DM$ interaction and external magnetic fields. Across all scenarios examined, coherence, as indicated by the negativity measure, exhibits temporal fluctuations, with these fluctuations becoming more pronounced in the presence of asymmetric external magnetic fields. Notably, the average coherence over time diminishes as the parameter $\alpha$ increases in both model settings. Furthermore, the augmentation of $D_z$ (representing $DM$ interaction strength) results in the convergence of the time-averaged coherence across all instances. The findings of this study are in alignment with prior research outcomes. The inherent symmetry of the problem ensures that all computations, visual representations, and interpretations concerning subsystem (qubit) $a$ are equally applicable to subsystem (qubit) $b$. In summary, exploring quantum entanglement in qubit systems holds significant promise for technology, security, and fundamental research.